\documentclass[fleqn]{article}
\usepackage[latin9]{inputenc}
\usepackage{geometry}
\usepackage{calc}
\usepackage{amsthm}
\usepackage{amsmath}
\usepackage{amssymb}
\usepackage{graphicx}
\usepackage{esint}

\makeatletter
  \theoremstyle{definition}
  \newtheorem{defn}{\protect\definitionname}
\theoremstyle{plain}
\newtheorem{thm}{\protect\theoremname}

\usepackage{amsfonts}

\makeatother

  \providecommand{\definitionname}{Definition}
\providecommand{\theoremname}{Theorem}

\begin{document}

\title{Review of Nelson's analysis of Bell's theorem}

\author{Benjamin Schulz, %
\thanks{Colmdorfstr. 32, 81249 München, e-mail: Benjamin.Schulz@physik.uni-muenchen.de%
} }

\maketitle
This article contains a review of Nelson's analysis of Bell's theorem. It shows that Bell's inequalities can be violated with a theory of local random variables if one accepts that the outcomes of these variables are not predetermined prior to measurement. The article describes the relation between Bell's theorem and the Strong Free Will theorem of Conway and Kochen. Then, the original articles of Bell are analyzed in detail. Following an article of Faris, it is explained that Bell's work on the hidden variable question in fact describes two separate theorems. Bell's first theorem says that there can be no model for the singlet state where an outcome does not depend locally on the settings of the detector where the outcome was measured. Bell's second theorem shows that Bell's inequalities can be violated by a theory that is either not deterministic, or violates causality in the sense of relativity or the free will assumption of the experimenters. It is shown in detail where Bell implicitly makes the various locality assumptions that Nelson has shown to be necessary for deriving Bell's inequality. The article closes by relating the various assumptions needed to derive Bell's theorem with the reality criterion of EPR

\tableofcontents{}

\section{Introduction}

In $1935$, Einstein, Podolsky and Rosen (EPR) wrote an article in
which they denied that quantum theory would be a complete theory of
nature \cite{Einstein}. Around $1951$, Bohm gave a more testable
outline of the so-called ``EPR paradox'' \cite{Bohm,Bohm2}. He
described a thought experiment with one source that ejects particles
having opposite spin to two spatially separate Stern-Gerlach magnets
of variable orientations, see Fig. \ref{fig:1}. Then, in $1964$,
Bell published a theorem about this paradox in the form of an inequality.
It made clear that hidden variable theories fulfilling certain conditions
would contradict quantum mechanics.

In his first publication on that topic, Bell wrote: ``If hidden parameters
would be added to quantum mechanics, there must be a mechanism, whereby
the setting of one measuring device can influence another spatially
separated device'' and the signal involved has to ``propagate instantaneously''
\cite{Bell2}. However, Bell's first contribution underwent several
modifications. Over the years, more and more instructive proofs of
his inequality were constructed by him and others. A collection of
all of Bell's fundamental articles can be found in \cite{Bell}. Finally
in $1969$, Clauser, Holt, Shimony and Horne (CHSH) brought the inequality
of Bell into a form suitable for experimental investigation \cite{HoltHorneClauserShimony}.

In $1985$, Nelson tried to analyse Bell's theorem with full mathematical
rigor. One result of his analysis was that Bell's work in \cite{Bell2}
and \cite{Bell3} basically described two separate theorems. According
to Faris \cite{Far}, Bell's earliest work \cite{Bell2} implies that
Bell's inequality can be derived without any hidden parameters or
locality assumption, if one just requires that the measurement devices
are unable to modify the outcomes and correlations at all 
(i.e if purely local influences of the devices are excluded, too). Bell's
later work from \cite{Bell3} uses a definition of ``local causality''
of a theory as a starting point to derive Bell's inequality. Nelson
found that this ``local causality'' assumption could be divided
into two separate conditions. Both conditions are necessary to derive
the inequality, but only one of them has to hold if in a stochastic
hidden variable theory an outcome depends on measurements made at
a spatially separated location. Nelson published his result twice
\cite{Ne3,Ne4}. (However, a small correction was added later \cite{Ne5},
in order to make Nelson's theorem compatible with Mermin's presentation
\cite{Mer1}.) Furthermore, there are studies from other mathematicians,
for example Faris \cite{Far}, which give further insightful analysis
of Nelson's work. Unfortunately, Nelson's articles on Bell's inequality
got almost overlooked by physicists. The reason for this might be
that Nelson found his own interpretation of quantum mechanics \cite{Ne1,Ne2}
to be at variance with the requirements of his theorem for a model
without instantaneous signaling effects.

This article is organized as follows: In section 2, we review Nelson's
contribution towards a mathematically rigorous understanding of Bell's
inequality. In section 3 we analyze the consequences of one assumption
that is necessary to derive Bell's theorem. In section 4, the relationship
between Bell's theorem and the so called Strong Free Will theorem
of Conway and Kochen\cite{Freewill} is elucidated. Section 5 analyzes
the original articles of Bell. Following an article of Faris \cite{Far},
we show that Bell's work in fact describes two separate theorems.
It is described where Bell implicitely makes the various locality
assumptions that Nelson has shown to be necessary for deriving Bell's
inequality. The article closes with section 6 by analyzing the relation
between the reality criteria of EPR and the locality assumptions that
are necessary to derive Bell's inequality. The article will need some
understanding of mathematical probability theory. For a general introduction
to this theory, see \cite{probab1,probab2,probab3,probab4,Far}.

\section{Nelson's analysis of Bell's theorem}

\subsection{The setup of the EPR experiment in theory}

\begin{figure}[t]
\begin{minipage}[t]{9.5cm}%
\fbox{\includegraphics[clip,scale=0.5985]{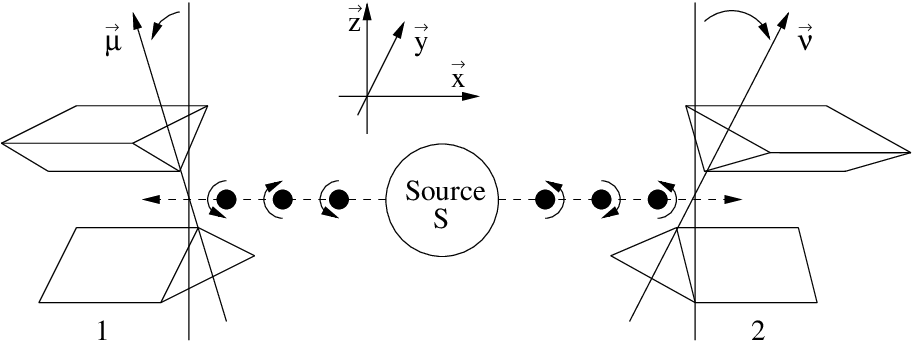}} \protect\caption{Drawing of an EPR experiment with Stern-Gerlach magnets $1$, $2$
and axes $\vec{\mu}$, $\vec{\nu}$. The black dots illustrate the
particles moving to the detectors. Their spin direction with respect
to the $\vec{z}$ axis is indicated by the small arrows around them.}

\label{fig:1} %
\end{minipage}\hfill{}%
\begin{minipage}[t]{5.25cm}%
\fbox{\includegraphics[clip,scale=0.4485]{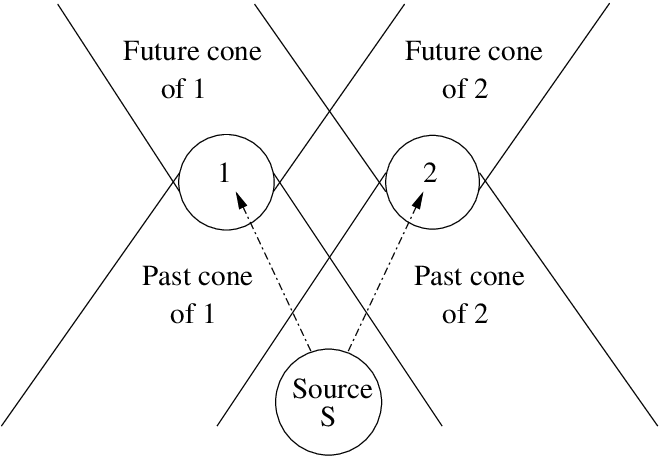}} \protect\caption{Space-time diagram of an EPR experiment. The dashed arrows display
particle trajectories}

\label{fig:2} %
\end{minipage}
\end{figure}

An EPR experiment consists of two measurement devices $1$ and $2$
which are space-like separated, as well as a particle source in the
intersection of their past cones. The source ejects pairs of particles
to the two detectors. At $1$ and $2$, the direction of the particle
spins is measured with a Stern-Gerlach magnet. These magnets can be
rotated around arbitrary directions. The direction of the magnets
is called $\vec{\mu}$ for detector $1$ and $\vec{\nu}$ for detector
$2$ (see Figs. \ref{fig:1}, \ref{fig:2}). 

When the particles arrive in the Stern-Gerlach magnets, the magnetic
field of these devices could change the particle's properties, including
the spin. In an EPR experiment, the parts of the detectors that might
influence the particles are the axes $\vec{\mu}$ and $\vec{\nu}$.
Those axes can be chosen by the experimenter freely at will. Due to
the Stern-Gerlach magnets, we have to describe our measurement results
by device dependent random variables. For the two detectors $1$ and
$2$, we define two axis dependent families of random variables $D_{1\vec{\mu}}:\Omega_{1D}\mapsto\Omega_{1D}'$
and $D_{2\vec{\nu}}:\Omega_{2D}\mapsto\Omega_{2D}'$ . The state spaces
of these random variables are $\Omega_{1D}'=\Omega_{2D}'=\left\{ \uparrow,\downarrow\right\} $
and $\Omega_{1D}$, $\Omega_{2D}$ are sample spaces. We denote the
outcomes of $D_{1\vec{\mu}}$ and $D_{2\vec{\nu}}$ on the state space
by the variables $\tilde{\sigma}_{1}$ and $\tilde{\sigma}_{2}$.

The results at the detectors could depend on the preparation of the
particles by the source. We denote the sample space of outcomes at
the source as $\Omega_{S}$ and say that the corresponding events
happen at preparation stage. Since we have a set of outcomes at the
source and at the detectors, we must describe the EPR experiment with
an enlarged probability space. We therefore define an enlarged sample
space $\Omega=\Omega_{1D}\times\Omega_{S}\times\Omega_{2D}$ . The
probability measure of the enlarged probability space will be denoted
by $\mathrm{P}$. 

On this enlarged probability space, we introduce a family of random
variables 
\begin{equation}
\phi_{1\vec{\mu}}\otimes\phi_{2\vec{\nu}}:\Omega\mapsto\Omega_{1D}'\times\Omega_{2D}'
\end{equation}
which is defined through the equation 
\begin{equation}
\left(\phi_{1\vec{\mu}}\otimes\phi_{2\vec{\nu}}\right)\left(\omega\right)=D_{1\vec{\mu}}\left(\omega_{1D}\right)\otimes D_{2\vec{\nu}}\left(\omega_{2D}\right)
\end{equation}
with $\omega\in\Omega$ and $\omega_{1D}\in\Omega_{1D}$, and $\omega_{2D}\in\Omega_{2D}$.

The random variable $\phi_{1\vec{\mu}}\otimes\phi_{2\vec{\nu}}$ gives
information on the joint outcomes that happen simultaneously at each
measurement station as a function of outcomes of the enlarged space.
To make contact with Nelson's notation, we define the notations $\sigma_{1}\equiv\tilde{\sigma}_{1}\times\Omega_{2D}'$,
$\sigma_{1}=\uparrow\equiv\uparrow\times\Omega_{2D}'$, and $\sigma_{1}=\downarrow\equiv\downarrow\times\Omega_{2D}'$.
Similarly, we define $\sigma_{2}\equiv\Omega_{1D}'\times\tilde{\sigma}_{2}$,
$\sigma_{2}=\uparrow\equiv\Omega_{1D}'\times\uparrow$, and $\sigma_{2}=\downarrow\equiv\Omega_{1D}'\times\downarrow$.

When we describe the EPR experiment, we often have to consider collections
of events. Given a power set $\mathcal{P}(\Omega)$, a sigma algebra
$\mathcal{F}$ is a subset $\mathcal{F\subseteq P}(\Omega)$, where
\begin{itemize}
\item $\Omega\in\mathcal{F}$
\item $A\in\mathcal{F}\Rightarrow A^{c}\in\mathcal{F}$ with $A^{c}$ being
the complement of $A$
\item $A_{1},A_{2},\ldots A_{n}\in\mathcal{F},n\in\mathbb{N}\Rightarrow\cup_{n\in\mathbb{N}}A_{n}\in\mathcal{F}$ 
\end{itemize}
Since $\cap_{n\in\mathbb{N}}A_{n}=\left(\cup_{n\in\mathbb{N}}A_{n}^{c}\right)^{c}\in\mathcal{F}$,
the intersection $\cap_{n\in\mathbb{N}}A_{n}$ is in $\mathcal{F}$,
too. This means that whenever events are combined by operations like
complement, union and intersection, the resulting events are still
in the sigma algebra, which is said to be closed under these operations. 

If a sigma algebra has only finitely many events, it is determined
by a partition, which is a collection of nonempty exclusive subsets
of the sample space $\Omega$, whose union is $\Omega$. The events
in the sigma algebra are then the sure event, the impossible event,
the events in the partition and the unions of the events in the partition.
We say the sigma algebra is generated by the events in the partition.
Faris \cite{Far} gives three illustrative examples for sigma algebras. 

The outcomes of an experiment where two coins are tossed that come
up heads H or tails T has a set of outcomes $\Omega=\left\{ HH,HT,TH,TT\right\} $.
The largest possible sigma algebra is then generated by the partition
into 4 events $\left\{ HH\right\} ,\left\{ HT\right\} ,\left\{ TH\right\} ,\left\{ TT\right\} $
and the sigma algebra consists of the sure event, the impossible event,
the events in the partition, and all possible unions of the events
in the partition. A smaller sigma algebra is able to specify the total
number of heads. This sigma algebra is generated by a partition into
three events $\left\{ HH\right\} ,\left\{ HT,TH\right\} ,\left\{ TT\right\} $.
Knowing which of these events happened implies that we know whether
the number of heads is $0,1$or $2$. The sigma algebra generated
by this partition then contains the impossible event, the sure event,
the events from the partition $\left\{ HH\right\} ,\left\{ HT,TH\right\} ,\left\{ TT\right\} $
and their unions $\left\{ HH,HT,TH\right\} ,\left\{ HH,TT\right\} ,\left\{ HT,TH,TT\right\} $.
Another example would be a sigma algebra that contains information
on whether the first toss is heads or tails. It contains the impossible
event, the sure event, and the events $\left\{ HH,HT\right\} $and
$\left\{ TH,TT\right\} $.

For the enlarged probability space of the EPR experiment with its
sample space $\Omega$, we define a large sigma algebra $\mathcal{F}=\mathcal{P}(\Omega)$
that contains all information of the experiment. The events $\left\{ \phi_{1\vec{\mu}}\otimes\phi_{2\vec{\nu}}\in\sigma_{1}=\downarrow\right\} $
and $\left\{ \phi_{1\vec{\mu}}\otimes\phi_{2\vec{\nu}}\in\sigma_{1}=\uparrow\right\} $are
defined on the enlarged probability space and only contain the information
on the outcomes at detector $1$ for an axis $\vec{\mu}$. We let
the partition of these events generate a sigma algebra $\mathcal{F}_{1}$.
This sigma algebra $\mathcal{F}_{1}$ then contains all information
about the events that might happen at $1$. Similarly, the events
$\left\{ \phi_{1\vec{\mu}}\otimes\phi_{2\vec{\nu}}\in\sigma_{2}=\downarrow\right\} $
and $\left\{ \phi_{1\vec{\mu}}\otimes\phi_{2\vec{\nu}}\in\sigma_{2}=\uparrow\right\} $
contain the information on the outcomes at detector $2$. The partition
of these events generates a sigma algebra denoted by $\mathcal{F}_{2}$
.

The probabilities of the events that are generated by $\phi_{1\vec{\mu}}\otimes\phi_{2\vec{\nu}}$
will be denoted by 
\begin{equation}
\mathrm{P}\left(\left\{ \phi_{1\vec{\mu}}\otimes\phi_{2\vec{\nu}}\in\left(\sigma_{1}\bigcap\sigma_{2}\right)\right\} \right)\equiv\mathrm{P}_{\phi_{1\vec{\mu}}\otimes\phi_{2\vec{\nu}}}\left(\sigma_{1}\bigcap\sigma_{2}\right).\label{eq:def123456}
\end{equation}
The expression $\mathrm{P}_{\phi_{1\vec{\mu}}\otimes\phi_{2\vec{\nu}}}$
defines some kind of axis dependent family of probability measures
for events of the form $\sigma_{1}\bigcap\sigma_{2}$.

Since the family of random variables $\phi_{1\vec{\mu}}\otimes\phi_{2\vec{\nu}}$
is defined on the enlarged probability space, it will give us the
possibility to investigate the relationship between the outcomes at
the detectors and the ones at preparation stage. We let a third sigma
algebra $\mathcal{F}_{S}$ contain the information about the outcomes
at the source. Accordingly, $\mathcal{F}_{S}$ is defined by the collection
of sets $\Omega_{1D}\times\mathcal{P}(\Omega_{S})\times\Omega_{2D}$,
where $\mathcal{P}$ denotes the power set.

We can compute the conditional probability of the events at the detectors
given the events in $\mathcal{F}_{S}$ that happen at preparation
stage with the conditional expectation value
\begin{equation}
\mathrm{P}\left(\left.A\right|\mathcal{F}_{S}\right)(\omega)\equiv\mathrm{EX}\left[\left.1_{A}\right|\mathcal{F}_{S}\right](\omega)\label{eq:specialeq}
\end{equation}
where $1_{A}(\omega)=\begin{cases}
1 & \forall\omega\in A\\
0 & \forall\omega\notin A
\end{cases}$ is the indicator random variable of $A$. Eq.\ (\ref{eq:specialeq})
is the definition of a random variable for outcomes $\omega$ in the
enlarged probability space. It can be interpreted as the revised probability
of an event $A$ to happen with respect to the extra information about
which events in $\mathcal{F}_{S}$ occur for a given outcome $\omega$.
According to Faris \cite{Far} and Bauer\cite{probab3}, $\mathrm{P}\left(\left.A\right|\mathcal{F}_{S}\right)$
fulfils the following properties: 
\begin{itemize}
\item For all events $A\in\mathcal{F}$, the random variable $\mathrm{P}\left(\left.A\right|\mathcal{F}_{S}\right)$
is measurable with respect to $\mathcal{F}_{S}$. 
\item We have $0\leq\mathrm{P}\left(\left.A\right|\mathcal{F}_{S}\right)\leq1$
for all events $A\in\mathcal{F}$ and outcomes on the enlarged probability
space. If $A$ is the sure event $\Omega_{1D}\times\Omega_{S}\times\Omega_{2D}$,
then $\mathrm{P}\left(\left.A\right|\mathcal{F}_{S}\right)=1$ for
all outcomes on the enlarged probability space. In case $A$ is the
impossible event $\varnothing$, then we get $\mathrm{P}\left(\left.A\right|\mathcal{F}_{S}\right)=0$
for all outcomes. 
\item For every sequence of pairwise disjoint events $\left(A_{n}\right)_{n\in\mathbb{N}}\in\mathcal{F}$,
the equality 
\[
\mathrm{P}\left(\left.\bigcup_{n=1}^{\infty}A_{n}\right|\mathcal{F}_{S}\right)=\sum_{n=1}^{\infty}\mathrm{P}\left(\left.A_{n}\right|\mathcal{F}_{S}\right)
\]
 holds. 
\item The unconditional probability of any event $A\in\mathcal{F}$ can
be computed with the expectation value 
\begin{equation}
\mathrm{EX}\left[\mathrm{P}\left(\left.A\right|\mathcal{F}_{S}\right)\right]=\int\mathrm{P}\left(\left.A\right|\mathcal{F}_{S}\right)\,\mathrm{dP}=\mathrm{P}\left(A\right)\label{eq:expectdsadf}
\end{equation}
. 
\item For any event $A\in\mathcal{F}$ and another event $A_{1S}\in\mathcal{F}_{S}$,
where $A_{1S}\neq\varnothing$, it follows from Eq.\  (\ref{eq:specialeq})
that $\mathrm{P}\left(\left.A\bigcap A_{1S}\right|\mathcal{F}_{S}\right)=\mathrm{P}\left(\left.A\right|\mathcal{F}_{S}\right)$
for outcomes in $A_{1S}$, and $\mathrm{P}\left(\left.A\bigcap A_{1S}\right|\mathcal{F}_{S}\right)=0$
for outcomes not in $A_{1S}$. 
\end{itemize}
In case that the events on the enlarged probability space are generated
by $\phi_{1\vec{\mu}}\otimes\phi_{2\vec{\nu}}$, we will use the following
notation: 
\begin{equation}
\mathrm{P}_{\phi_{1\vec{\mu}}\otimes\phi_{2\vec{\nu}}}\left(\left.\sigma_{1}\cap\sigma_{2}\right|\mathcal{F}_{S}\right)\equiv\mathrm{P}\left(\left\{ \left.\phi_{1\vec{\mu}}\otimes\phi_{2\vec{\nu}}\in\left(\sigma_{1}\cap\sigma_{2}\right)\right\} \right|\mathcal{F}_{S}\right).\label{eq:def}
\end{equation}

In the analysis of the EPR experiment, we will often have to deal
with equivalent events. Two events $A$ and $B$ in $\mathcal{F}$
are equivalent if

\begin{equation}
\mathrm{P}\left(A\bigcap B\right)=\mathrm{P}(A)=\mathrm{P}(B).
\end{equation}
This means that if $A$ happens, then $B$ also happens and vice versa.
This does not mean, however, that $A$ causes $B$ to happen or vice-versa.
If $A$ and $B$ are equivalent to each other, then their complements
$(A)^{C}$ and $(B)^{C}$ must also be equivalent to each other. Furthermore,
one can show that if the event $A$ is equivalent to the event $B$,
and $B$ is equivalent to another event $C$, then $A$ is also equivalent
to $C$. 

We also need a certain definition of equivalent random variables.
We say that two random variables $\phi_{A}$ and $\phi_{B}$ are equivalent
if 
\[
\phi_{A}=\phi_{B}
\]

Now, all the mathematical structures needed to analyse the EPR experiment
have been defined and we can proceed with the necessary locality conditions.

\subsection{Active locality}

At first, Nelson defined two different forms of locality: Active locality
and passive locality. The meaning of active locality is: Whatever
axes the experimenter selects at one measurement device, e.\,g. at
$2$, it does not change the outcomes at $1$, as long as $1$ does
not lie in the future cone of $2$. Active locality therefore contains
the assumption that the experimenters can choose the axes of their
measurement instruments freely, and it contains the causality requirement
from the theory of relativity. We can mathematically define active
locality as follows:
\begin{defn}
The Stern Gerlach magnets may influence the outcomes of the experiment
as a function of their axes. The corresponding events are generated
by the axis dependent random variables $\phi_{1\vec{\mu}}\otimes\phi_{2\vec{\nu}}$,
where $\phi_{1\vec{\mu}}$ generates the outcome at detector 1 with
axis $\vec{\mu}$ and $\phi_{2\vec{\nu}}$ generates the outcome at
detector 2 with axis $\vec{\nu}$. Now we let the area of measurement
station $1$ be disjoint from the future cone of station $2$. With
the axis $\vec{\mu}$ of $1$ being left constant, different axes
$\vec{\nu}\neq\vec{\nu}'$ of Stern-Gerlach magnet $2$ are chosen.
A choice of an axis at 2 may influence the outcome at 1. We denote
the random variable at 1 that generates the possibly changed outcome
as $\hat{\phi}_{1\vec{\mu}}.$We call a theory actively local if for
events $\left\{ \phi_{1\vec{\mu}}\otimes\phi_{2\vec{\nu}}\in\sigma_{1}\right\} \in\mathcal{F}_{1}$
and $\left\{ \hat{\phi}_{1\vec{\mu}}\otimes\phi_{2\vec{\nu}'}\in\sigma_{1}\right\} \in\mathcal{F}_{1}$,
the random variables $\phi_{1\vec{\mu}}$ and $\hat{\phi}_{1\vec{\mu}}$
are equivalent, or 
\begin{equation}
\phi_{1\vec{\mu}}=\hat{\phi}_{1\vec{\mu}}.\label{eq:activeloc}
\end{equation}
This guarantees that the outcomes at 1 remains the same regardeless
of the axes that are chosen at 2, and we get for the probabilities
of these events:

\begin{equation}
\mathrm{P}_{\hat{\phi}_{1\vec{\mu}}\otimes\phi_{2\vec{\nu}'}}\left(\sigma_{1}\right)=\mathrm{P}_{\phi_{1\vec{\mu}}\otimes\phi_{2\vec{\nu}'}}\left(\sigma_{1}\right)=\mathrm{P}_{\phi_{1\vec{\mu}}\otimes\phi_{2\vec{\nu}}}\left(\sigma_{1}\right).\label{eq:activeloca}
\end{equation}
Note that the probabilities of Eq. (\ref{eq:activeloca}) are actually
required by the quantum mechanics, whereas Eq. (\ref{eq:activeloc})
is a statement on the outcomes that is not required by the quantum
mechanical formalism.

Similarly, in an actively local theory, the random variables $\phi_{2\vec{\nu}}$
of an event $\left\{ \phi_{1\vec{\mu}}\otimes\phi_{2\vec{\nu}}\in\sigma_{2}\right\} \in\mathcal{F}_{2}$
and $\hat{\phi}_{2\vec{\nu}}$ from $\left\{ \phi_{1\vec{\mu}'}\otimes\hat{\phi}_{2\vec{\nu}}\in\sigma_{2}\right\} \in\mathcal{F}_{2}$
with $\mu'\neq\mu$ should be equivalent, 
\begin{equation}
\hat{\phi}_{2\vec{\nu}}=\phi_{2\vec{\nu}}\label{eq:activeloc2-1}
\end{equation}
 provided that the spatial region of the measurement station $2$
is disjoint from the future cone of the magnet at $1$. Since by Eq.
(\ref{eq:activeloc2-1}) we have the same outcomes at 2 for the different
axes at 1, we get for the probabilities:

\begin{eqnarray}
\mathrm{P}_{\phi_{1\vec{\mu}'}\otimes\hat{\phi}_{2\vec{\nu}}}\left(\sigma_{2}\right) & =\mathrm{P}_{\phi_{1\vec{\mu}'}\otimes\phi_{2\vec{\nu}}}\left(\sigma_{2}\right)= & \mathrm{P}_{\phi_{1\vec{\mu}}\otimes\phi_{2\vec{\nu}}}\left(\sigma_{2}\right).\label{eq:activeloc2}
\end{eqnarray}

\end{defn}
By Eq.\ (\ref{eq:def}), we can condition the probabilities of $\left\{ \phi_{1\vec{\mu}}\otimes\phi_{2\vec{\nu}}\in\sigma_{1}\right\} \in\mathcal{F}_{1}$
and $\left\{ \phi_{1\vec{\mu}}\otimes\phi_{2\vec{\nu}}\in\sigma_{2}\right\} \in\mathcal{F}_{2}$
with respect to $\mathcal{F}_{S}$ and we get 
\begin{eqnarray}
\mathrm{P}_{\phi_{1\vec{\mu}}\otimes\phi_{2\vec{\nu}}}\left(\left.\sigma_{1}\right|\mathcal{F}_{S}\right) & = & \mathrm{P}_{\phi_{1\vec{\mu}}\otimes\phi_{2\vec{\nu}'}}\left(\left.\sigma_{1}\right|\mathcal{F}_{S}\right),\nonumber \\
\mathrm{P}_{\phi_{1\vec{\mu}}\otimes\phi_{2\vec{\nu}}}\left(\left.\sigma_{2}\right|\mathcal{F}_{S}\right) & = & \mathrm{P}_{\phi_{1\vec{\mu}'}\otimes\phi_{2\vec{\nu}}}\left(\left.\sigma_{2}\right|\mathcal{F}_{S}\right).\label{eq:activeloc3}
\end{eqnarray}

It is important to note that Eqs. (\ref{eq:activeloca}) and (\ref{eq:activeloc2})
describe probabilities which are given by quantum mechanics. These
equations forbids in any theory that describes the quantum mechanical
probabilities to send instantaneous signals to space-like separated
locations. In contrast Eqs. (\ref{eq:activeloc}) and (\ref{eq:activeloc2-1})
are not given by quantum mechanics and are violated by many hidden
variable theories, such as Bohmian mechanics, where the measurement
of individual outcomes at one station can influence another outcome
at a separated station. 

Because quantum mechanics does not contain Eqs. (\ref{eq:activeloc})
and (\ref{eq:activeloc2-1}), it does not forbid an explanation of
its outcomes with a model that violates causality in the sense of
special relativity. Someone who strictly insists on the relativity
principle could therefore view quantum mechanics as an incomplete
theory.

One should note that for an event $\left\{ \phi_{1\vec{\mu}}\otimes\phi_{2\vec{\nu}}\in\left(\sigma_{1}\bigcap\sigma_{2}\right)\right\} $,
which is neither in $\mathcal{F}_{1}$ or $\mathcal{F}_{2}$, one
can not conclude from the above active locality definitions that we
would have for some pairs of axes, where $\mu'\neq\mu$ and $\nu'\neq\nu$,
and equation like 
\[
\phi_{1\vec{\mu}}\otimes\phi_{2\vec{\nu}}=\phi_{1\vec{\mu}'}\otimes\phi_{2\vec{\nu}'}
\]
since the random variables $\phi_{1\vec{\mu}}$ and $\phi_{1\vec{\mu}'}$
do not have to be equivalent. Similarly, the random variables $\phi_{1\vec{\mu}}$
and $\phi_{1\vec{\mu}'}$do not have to be equivalent for $\nu'\neq\nu$.
Hence 
\[
\phi_{1\vec{\mu}}\otimes\phi_{2\vec{\nu}}\neq\phi_{1\vec{\mu}'}\otimes\phi_{2\vec{\nu}'}
\]
The physical reason for this is that spin operators, like all angular
momentum operators in quantum mechanics, do not commute.

\subsection{Passive locality}
\begin{defn}
We consider the conditional joint probability of $\left\{ \phi_{1\vec{\mu}}\otimes\phi_{2\vec{\nu}}\in\sigma_{1}\right\} \in\mathcal{F}_{1}$
and $\left\{ \phi_{1\vec{\mu}}\otimes\phi_{2\vec{\nu}}\in\sigma_{2}\right\} \in\mathcal{F}_{2}$
with respect to $\mathcal{F}_{S}$. It gives information about events
which happen simultaneously at the spatially separated locations $1$
and $2$. We say that passive locality holds if 
\begin{equation}
\mathrm{P}_{\phi_{1\vec{\mu}}\otimes\phi_{2\vec{\nu}}}\left(\left.\sigma_{1}\bigcap\sigma_{2}\right|\mathcal{F}_{S}\right)=\mathrm{P}_{\phi_{1\vec{\mu}}\otimes\phi_{2\vec{\nu}}}\left(\left.\sigma_{1}\right|\mathcal{F}_{S}\right)\mathrm{P}_{\phi_{1\vec{\mu}}\otimes\phi_{2\vec{\nu}}}\left(\left.\sigma_{2}\right|\mathcal{F}_{S}\right),\label{eq:passiveloc}
\end{equation}
for every pair of axes $\vec{\mu}$ and $\vec{\nu}$. 
\end{defn}
The outcomes of an experiment may be statistically dependent. Such
a dependence may arise because of a preparation stage. The condition
of passive locality says that the outcomes should be conditionally
independent given the information about the events at preparation
stage which are $\mathcal{F}_{S}$. A violation of passive locality
would mean that the dependence of the outcomes does not originate
at $\mathcal{F}_{S}$. It is possible to have active locality without
passive locality. Furthermore, Nelson writes that a theory which violates
passive locality does not have to incorporate any non-local interaction
between the spatially separated measurement stations $1$ and $2$.

\subsection{Bell's second theorem (in Nelson's notation)}
\begin{thm}
(Bell's second theorem, Bell, Nelson): If active and passive locality
hold and
\begin{equation}
\mathrm{P}_{\phi_{1\vec{\mu}}\otimes\phi_{2\vec{\mu}}}\left(\sigma_{1}=\uparrow\bigcap\sigma_{2}=\downarrow\right)+\mathrm{P}_{\phi_{1\vec{\mu}}\otimes\phi_{2\vec{\mu}}}\left(\sigma_{1}=\downarrow\bigcap\sigma_{2}=\uparrow\right)=1,\label{eq:1}
\end{equation}
then 
\begin{equation}
\left|\mathrm{E}\left(\vec{\mu},\vec{\nu}\right)-\mathrm{E}\left(\vec{\mu},\vec{\nu}'\right)+\mathrm{E}\left(\vec{\mu}',\vec{\nu}\right)-\mathrm{E}\left(\vec{\mu}',\vec{\nu}'\right)\right|\leq2.\label{eq:2}
\end{equation}
where the function $\mathrm{E}(\vec{\mu},\vec{\nu})$ is called correlation
coefficient and is defined by
\begin{align}
\mathrm{E}\left(\vec{\mu},\vec{\nu}\right) & \equiv\mathrm{P}_{\phi_{1\vec{\mu}}\otimes\phi_{2\vec{\nu}}}\left(\sigma_{1}=\uparrow\bigcap\sigma_{2}=\uparrow\right)+\mathrm{P}_{\phi_{1\vec{\mu}}\otimes\phi_{2\vec{\nu}}}\left(\sigma_{1}=\downarrow\bigcap\sigma_{2}=\downarrow\right)\nonumber \\
 & \quad-\mathrm{P}_{\phi_{1\vec{\mu}}\otimes\phi_{2\vec{\nu}}}\left(\sigma_{1}=\uparrow\bigcap\sigma_{2}=\downarrow\right)-\mathrm{P}_{\phi_{1\vec{\mu}}\otimes\phi_{2\vec{\nu}}}\left(\sigma_{1}=\downarrow\bigcap\sigma_{2}=\uparrow\right).\label{eq:2a}
\end{align}

\end{thm}
Note that quantum mechanics fulfillfs Eq. (\ref{eq:1}) but violates
the inequality of Eq. (\ref{eq:2}). So for any theory that reproduces
the quantum mechanical probabilities, either active or passive locality
has to fail. Nelson's proof \cite{Ne3,Ne4} (with corrections in \cite{Ne5})
proceeds as follows:
\begin{proof}
Equation (\ref{eq:1}) implies that if the axes of the Stern-Gerlach
magnets are the same, the spin values measured at $1$ and $2$ are
always opposite. Using 
\[
\mathrm{P}_{\phi_{1\vec{\mu}}\otimes\phi_{2\vec{\mu}}}\left(\sigma_{1}=\uparrow\right)=\mathrm{P}_{\phi_{1\vec{\mu}}\otimes\phi_{2\vec{\mu}}}\left(\sigma_{1}=\downarrow\right)=\frac{1}{2},
\]
\[
\mathrm{P}_{\phi_{1\vec{\mu}}\otimes\phi_{2\vec{\mu}}}\left(\sigma_{2}=\uparrow\right)=\mathrm{P}_{\phi_{1\vec{\mu}}\otimes\phi_{2\vec{\mu}}}\left(\sigma_{2}=\downarrow\right)=\frac{1}{2},
\]
 and 
\[
\mathrm{P}_{\phi_{1\vec{\mu}}\otimes\phi_{2\vec{\mu}}}\left(\sigma_{1}=\uparrow\bigcap\sigma_{2}=\downarrow\right)=\mathrm{P}_{\phi_{1\vec{\mu}}\otimes\phi_{2\vec{\mu}}}\left(\sigma_{1}=\downarrow\bigcap\sigma_{2}=\uparrow\right)
\]
, we can rewrite Eq.(\ref{eq:1}) in the following form: 
\begin{equation}
\left\{ \begin{array}{ccccc}
\mathrm{P}_{\phi_{1\vec{\mu}}\otimes\phi_{2\vec{\mu}}}\left(\sigma_{1}=\uparrow\right) & = & \mathrm{P}_{\phi_{1\vec{\mu}}\otimes\phi_{2\vec{\mu}}}\left(\sigma_{1}=\uparrow\bigcap\sigma_{2}=\downarrow\right) & = & \mathrm{P}_{\phi_{1\vec{\mu}}\otimes\phi_{2\vec{\mu}}}\left(\sigma_{2}=\downarrow\right),\\
+ &  & + &  & +\\
\mathrm{P}_{\phi_{1\vec{\mu}}\otimes\phi_{2\vec{\mu}}}\left(\sigma_{1}=\downarrow\right) & = & \mathrm{P}_{\phi_{1\vec{\mu}}\otimes\phi_{2\vec{\mu}}}\left(\sigma_{1}=\downarrow\bigcap\sigma_{2}=\uparrow\right) & = & \mathrm{P}_{\phi_{1\vec{\mu}}\otimes\phi_{2\vec{\mu}}}\left(\sigma_{2}=\uparrow\right).
\end{array}\right.\label{eq:1abcdefg}
\end{equation}
This makes clear that the events at the detectors are equivalent if
the same axes at the separated measurement devices were chosen.

We are interested in the properties of the conditional probabilities
of the events $\left\{ \phi_{1\vec{\mu}}\otimes\phi_{2\vec{\nu}}\in\sigma_{1}\right\} \in\mathcal{F}_{1}$
and $\left\{ \phi_{1\vec{\mu}}\otimes\phi_{2\vec{\nu}}\in\sigma_{2}\right\} \in\mathcal{F}_{2}$
given the sigma algebra $\mathcal{F}_{S}$. At first, we will investigate,
what the equivalence property of Eq. (\ref{eq:1abcdefg}) combined
with the assumption of passive locality implies for these conditional
probabilities.

If the axes at the two detectors are the same, i.\ e. $\vec{\mu}=\vec{\nu}$,
we have, due to passive locality:
\begin{equation}
\mathrm{P}_{\phi_{1\vec{\mu}}\otimes\phi_{2\vec{\mu}}}\left(\left.\sigma_{1}\bigcap\sigma_{2}\right|\mathcal{F}_{S}\right)=\mathrm{P}_{\phi_{1\vec{\mu}}\otimes\phi_{2\vec{\mu}}}\left(\left.\sigma_{1}\right|\mathcal{F}_{S}\right)\mathrm{P}_{\phi_{1\vec{\mu}}\otimes\phi_{2\vec{\mu}}}\left(\left.\sigma_{2}\right|\mathcal{F}_{S}\right).\label{eq:a}
\end{equation}
Since $0\leq\mathrm{P}_{\phi_{1\vec{\mu}}\otimes\phi_{2\vec{\mu}}}\left(\left.\sigma_{2}\right|\mathcal{F}_{S}\right)\leq1$,
we get 
\begin{equation}
\mathrm{P}_{\phi_{1\vec{\mu}}\otimes\phi_{2\vec{\mu}}}\left(\left.\sigma_{1}\bigcap\sigma_{2}\right|\mathcal{F}_{S}\right)\leq\mathrm{P}_{\phi_{1\vec{\mu}}\otimes\phi_{2\vec{\mu}}}\left(\left.\sigma_{1}\right|\mathcal{F}_{S}\right).\label{eq:ba}
\end{equation}
With the properties of the conditional probabilities stated in section
$2.1$, it follows that 
\begin{equation}
\mathrm{P}_{\phi_{1\vec{\mu}}\otimes\phi_{2\vec{\mu}}}\left(\sigma_{1}\right)=\mathrm{EX}\left[\mathrm{P}_{\phi_{1\vec{\mu}}\otimes\phi_{2\vec{\mu}}}\left(\left.\sigma_{1}\right|\mathcal{F}_{S}\right)\right]\label{eq:gga}
\end{equation}
and, similarly, 
\begin{equation}
\mathrm{P}_{\phi_{1\vec{\mu}}\otimes\phi_{2\vec{\mu}}}\left(\sigma_{1}\bigcap\sigma_{2}\right)=\mathrm{EX}\left[\mathrm{P}_{\phi_{1\vec{\mu}}\otimes\phi_{2\vec{\mu}}}\left(\left.\sigma_{1}\bigcap\sigma_{2}\right|\mathcal{F}_{S}\right)\right].\label{eq:ca}
\end{equation}
Using Eq. (\ref{eq:1abcdefg}), we observe that an event $\left\{ \phi_{1\vec{\mu}}\otimes\phi_{2\vec{\nu}}\in\sigma_{1}\right\} $
at detector $1$ implies an equivalent event $\left\{ \phi_{1\vec{\mu}}\otimes\phi_{2\vec{\nu}}\in\sigma_{2}\right\} $
at $2$, i.\ e.
\begin{equation}
\mathrm{P}_{\phi_{1\vec{\mu}}\otimes\phi_{2\vec{\mu}}}\left(\sigma_{1}\right)=\mathrm{P}_{\phi_{1\vec{\mu}}\otimes\phi_{2\vec{\mu}}}\left(\sigma_{1}\bigcap\sigma_{2}\right)=\mathrm{P}_{\phi_{1\vec{\mu}}\otimes\phi_{2\vec{\mu}}}\left(\sigma_{2}\right).\label{eq:d}
\end{equation}
Hence, the expectation values in Eqs. (\ref{eq:gga}) and (\ref{eq:ca})
are equal. Accordingly, we get with Eq. (\ref{eq:ba}): 
\begin{equation}
\mathrm{P}_{\phi_{1\vec{\mu}}\otimes\phi_{2\vec{\mu}}}\left(\left.\sigma_{1}\bigcap\sigma_{2}\right|\mathcal{F}_{S}\right)=\mathrm{P}_{\phi_{1\vec{\mu}}\otimes\phi_{2\vec{\mu}}}\left(\left.\sigma_{1}\right|\mathcal{F}_{S}\right),\label{eq:ea}
\end{equation}
and 
\begin{equation}
\mathrm{P}_{\phi_{1\vec{\mu}}\otimes\phi_{2\vec{\mu}}}\left(\left.\sigma_{1}\bigcap\sigma_{2}\right|\mathcal{F}_{S}\right)=\mathrm{P}_{\phi_{1\vec{\mu}}\otimes\phi_{2\vec{\mu}}}\left(\left.\sigma_{2}\right|\mathcal{F}_{S}\right).\label{eq:fa}
\end{equation}

Now, we will look at the consequences of active and passive locality
for the conditional probabilities. Passive locality demands for events
$\left\{ \phi_{1\vec{\mu}}\otimes\phi_{2\vec{\nu}}\in\sigma_{1}=\uparrow\right\} $
at $1$ and $\left\{ \phi_{1\vec{\mu}}\otimes\phi_{2\vec{\nu}}\in\sigma_{2}=\downarrow\right\} $
at $2$ that:
\begin{eqnarray}
\mathrm{P}_{\phi_{1\vec{\mu}}\otimes\phi_{2\vec{\nu}}}\left(\left.\sigma_{1}=\uparrow\bigcap\sigma_{2}=\downarrow\right|\mathcal{F}_{S}\right) & = & \mathrm{P}_{\phi_{1\vec{\mu}}\otimes\phi_{2\vec{\nu}}}\left(\left.\sigma_{1}=\uparrow\right|\mathcal{F}_{S}\right)\nonumber \\
 &  & \times\mathrm{P}_{\phi_{1\vec{\mu}}\otimes\phi_{2\vec{\nu}}}\left(\left.\sigma_{2}=\downarrow\right|\mathcal{F}_{S}\right).\label{eq:4}
\end{eqnarray}
By Eq. (\ref{eq:activeloc3}), active locality implies that 
\begin{eqnarray}
\mathrm{P}_{\phi_{1\vec{\mu}}\otimes\phi_{2\vec{\nu}}}\left(\left.\sigma_{1}=\uparrow\right|\mathcal{F}_{S}\right) & = & \mathrm{P}_{\phi_{1\vec{\mu}}\otimes\phi_{2\vec{\mu}}}\left(\left.\sigma_{1}=\uparrow\right|\mathcal{F}_{S}\right)\equiv P_{\vec{\mu}}.\label{eq:6}
\end{eqnarray}
and due to Eq. (\ref{eq:activeloc3}), an analogous expression is
true for the events at $2$: 
\begin{equation}
\mathrm{P}_{\phi_{1\vec{\mu}}\otimes\phi_{2\vec{\nu}}}\left(\left.\sigma_{2}=\downarrow\right|\mathcal{F}_{S}\right)=\mathrm{P}_{\phi_{1\vec{\nu}}\otimes\phi_{2\vec{\nu}}}\left(\left.\sigma_{2}=\downarrow\right|\mathcal{F}_{S}\right).\label{eq:7}
\end{equation}

Without loss of generality, we can select an axis $\vec{\mu}=\vec{\nu}$
with $\vec{\nu}$ from Eq.\ (\ref{eq:7}) in the Eqs.\ (\ref{eq:1abcdefg}),
(\ref{eq:ea}) and (\ref{eq:fa}). It then follows from Eq.\ (\ref{eq:1abcdefg})
that the event $\left\{ \phi_{1\vec{\nu}}\otimes\phi_{2\vec{\nu}}\in\sigma_{1}=\downarrow\right\} $
at detector $2$ in the right hand side of Eq.\ (\ref{eq:7}) implies
an equivalent event of $\left\{ \phi_{1\vec{\nu}}\otimes\phi_{2\vec{\nu}}\in\sigma_{1}=\uparrow\right\} $
at $1$. Using the Eqs.\ (\ref{eq:ea}) and (\ref{eq:fa}) with an
axis $\vec{\mu}=\vec{\nu}$, we can conclude that 
\begin{eqnarray}
\mathrm{P}_{\phi_{1\vec{\mu}}\otimes\phi_{2\vec{\nu}}}\left(\left.\sigma_{2}=\downarrow\right|\mathcal{F}_{S}\right) & = & \mathrm{P}_{\phi_{1\vec{\nu}}\otimes\phi_{2\vec{\nu}}}\left(\left.\sigma_{2}=\downarrow\right|\mathcal{F}_{S}\right)\nonumber \\
 & = & \mathrm{P}_{\phi_{1\vec{\nu}}\otimes\phi_{2\vec{\nu}}}\left(\left.\sigma_{1}=\uparrow\right|\mathcal{F}_{S}\right)\equiv P_{\vec{\nu}}.\label{eq:10}
\end{eqnarray}
Plugging Eq.\ (\ref{eq:6}) and Eq.\ (\ref{eq:10}) back to Eq.\ (\ref{eq:4}),
we arrive at 
\begin{eqnarray}
\mathrm{P}_{\phi_{1\vec{\mu}}\otimes\phi_{2\vec{\nu}}}\left(\left.\sigma_{1}=\uparrow\bigcap\sigma_{2}=\downarrow\right|\mathcal{F}_{S}\right) & = & P_{\vec{\mu}}P_{\vec{\nu}}.\label{eq:11}
\end{eqnarray}

The events $\left\{ \phi_{1\vec{\mu}}\otimes\phi_{2\vec{\nu}}\in\sigma_{2}=\uparrow\right\} \in\mathcal{F}_{2}$
and $\left\{ \phi_{1\vec{\mu}}\otimes\phi_{2\vec{\nu}}\in\sigma_{2}=\downarrow\right\} \in\mathcal{F}_{2}$
are disjoint and their union is the sure event. According to section
$2.1$, the sum of the conditional probabilities with respect to a
sigma algebra is equal to unity for such events. Therefore, we get
with Eq.\ (\ref{eq:10}):
\begin{equation}
\mathrm{P}_{\phi_{1\vec{\mu}}\otimes\phi_{2\vec{\nu}}}\left(\left.\sigma_{2}=\uparrow\right|\mathcal{F}_{S}\right)=\mathrm{P}_{\phi_{1\vec{\nu}}\otimes\phi_{2\vec{\nu}}}\left(\left.\sigma_{2}=\uparrow\right|\mathcal{F}_{S}\right)=1-P_{\vec{\nu}}.\label{eq:17a}
\end{equation}
Similarly, 
\begin{eqnarray}
\mathrm{P}_{\phi_{1\vec{\mu}}\otimes\phi_{2\vec{\nu}}}\left(\left.\sigma_{1}=\downarrow\right|\mathcal{F}_{S}\right) & = & \mathrm{P}_{\phi_{1\vec{\mu}}\otimes\phi_{2\vec{\mu}}}\left(\left.\sigma_{1}=\downarrow\right|\mathcal{F}_{S}\right)\nonumber \\
 & = & 1-\mathrm{P}_{\phi_{1\vec{\mu}}\otimes\phi_{2\vec{\mu}}}\left(\left.\sigma_{1}=\uparrow\right|\mathcal{F}_{S}\right)=1-P_{\vec{\mu}}.\label{eq:20}
\end{eqnarray}
Due to passive locality and the Eqs.\ (\ref{eq:20}) and (\ref{eq:17a}),
we have for the events $\left\{ \phi_{1\vec{\mu}}\otimes\phi_{2\vec{\nu}}\in\sigma_{1}=\downarrow\right\} $
and $\left\{ \phi_{1\vec{\mu}}\otimes\phi_{2\vec{\nu}}\in\sigma_{2}=\uparrow\right\} $
:
\begin{eqnarray}
\mathrm{P}_{\phi_{1\vec{\mu}}\otimes\phi_{2\vec{\nu}}}\left(\left.\sigma_{1}=\downarrow\bigcap\sigma_{2}=\uparrow\right|\mathcal{F}_{S}\right) & = & \left(1-P_{\vec{\mu}}\right)\left(1-P_{\vec{\nu}}\right).\label{eq:21}
\end{eqnarray}
In the same way, we can derive the relations
\begin{equation}
\mathrm{P}_{\phi_{1\vec{\mu}}\otimes\phi_{2\vec{\nu}}}\left(\left.\sigma_{1}=\uparrow\bigcap\sigma_{2}=\uparrow\right|\mathcal{F}_{S}\right)=P_{\vec{\mu}}\left(1-P_{\vec{\nu}}\right)\label{eq:22}
\end{equation}
and 
\begin{equation}
\mathrm{P}_{\phi_{1\vec{\mu}}\otimes\phi_{2\vec{\nu}}}\left(\left.\sigma_{1}=\downarrow\bigcap\sigma_{2}=\downarrow\right|\mathcal{F}_{S}\right)=\left(1-P_{\vec{\mu}}\right)P_{\vec{\nu}}.\label{eq:23}
\end{equation}

Using the Eqs.\ (\ref{eq:22}), (\ref{eq:23}), (\ref{eq:11}) and
(\ref{eq:21}), we may define the function
\begin{eqnarray}
\mathrm{E}\left(\left.\vec{\mu},\vec{\nu}\right|\mathcal{F}_{S}\right) & \equiv & P_{\vec{\mu}}\left(1-P_{\vec{\nu}}\right)+\left(1-P_{\vec{\mu}}\right)P_{\vec{\nu}}-P_{\vec{\mu}}P_{\vec{\nu}}-\left(1-P_{\vec{\mu}}\right)\left(1-P_{\vec{\nu}}\right).\label{eq:24}
\end{eqnarray}
The conditional probabilities in Eq.\ (\ref{eq:24}) are all in the
range $0\leq\mathrm{P}_{\phi_{1\vec{\mu}}\otimes\phi_{2\vec{\nu}}}\left(\left.\sigma_{1\mu}\bigcap\sigma_{2\nu}\right|\mathcal{F}_{S}\right)\leq1$.
Therefore, the following inequality can be computed with four arbitrary
axes $\vec{\mu},\vec{\mu}'$ and $\vec{\nu},\vec{\nu}'$ at the two
stations $1$ and $2$: 
\begin{equation}
\left|\mathrm{E}\left(\left.\vec{\mu},\vec{\nu}\right|\mathcal{F}_{S}\right)+\mathrm{E}\left(\left.\vec{\mu},\vec{\nu}'\right|\mathcal{F}_{S}\right)+\mathrm{E}\left(\left.\vec{\mu}',\vec{\nu}\right|\mathcal{F}_{S}\right)-\mathrm{E}\left(\left.\vec{\mu}',\vec{\nu}'\right|\mathcal{F}_{S}\right)\right|\leq2.\label{eq:25}
\end{equation}
If we replace the conditional probabilities in Eq.\ (\ref{eq:25})
by their corresponding unconditional probability measures, we arrive
at the correlation coefficient from Eq.\ (\ref{eq:2}). The unconditional
probabilities are also in the range of $0\leq\mathrm{P}_{\phi_{1\vec{\mu}}\otimes\phi_{2\vec{\nu}}}\left(\sigma_{1\mu}\bigcap\sigma_{2\nu}\right)\leq1$,
and they are given by the expectation values of the conditional probabilities.
Hence, an inequality analogous to Eq. (\ref{eq:25}) must be true
for them: 
\begin{equation}
\left|\mathrm{E}\left(\vec{\mu},\vec{\nu}\right)+\mathrm{E}\left(\vec{\mu},\vec{\nu}'\right)+\mathrm{E}\left(\vec{\mu}',\vec{\nu}\right)-\mathrm{E}\left(\vec{\mu}',\vec{\nu}'\right)\right|\leq2.\label{eq:26}
\end{equation}

\end{proof}
Equation (\ref{eq:26}) is called Clauser-Holt-Shimony-Horne inequality
\cite{HoltHorneClauserShimony}. It is a version of Bell's inequality
\begin{equation}
\left|\mathrm{E}\left(\vec{\mu},\vec{\nu}\right)-\mathrm{E}\left(\vec{\mu},\vec{\nu}'\right)\right|-\mathrm{E}\left(\vec{\nu},\vec{\nu}'\right)\leq1,\label{eq:Bellorig}
\end{equation}
which can be similarly derived. Both Eq.\ (\ref{eq:26}) and Eq.\ (\ref{eq:Bellorig})
are violated in quantum mechanics and this violation was confirmed
experimentally in $1982$\cite{Aspect}. It should be noted that Jarret
\cite{Jar} arrived at a similar conclusion even though he did not
formulate his article within rigorous probability theory.

\section{Implications of passive locality}

We begin by analyzing the consequences of passive locality. Faris
showed in \cite{Far} that passive locality, when combined with relation
(\ref{eq:1}) from quantum mechanics, immediately leads to another
condition, which he calls ``deterministic passive locality''. According
to Faris, a theory fulfills the deterministic passive locality condition
if the following theorem holds:
\begin{thm}
Let the events $\left\{ \phi_{1\vec{\mu}}\otimes\phi_{2\vec{\mu}}\in\sigma_{1}\right\} \in\mathcal{F}_{1}$and
$\left\{ \phi_{1\vec{\mu}}\otimes\phi_{2\vec{\mu}}\in\sigma_{2}\right\} \in\mathcal{F}_{2}$
be equivalent with respect to $\mathrm{P}_{\phi_{1\vec{\mu}}\otimes\phi_{2\vec{\mu}}}$
and passive locality hold. Then, there must be an event $\mu_{1S}\in\mathcal{F}_{S}$
at preparation stage, which is equivalent to both $\left\{ \phi_{1\vec{\mu}}\otimes\phi_{2\vec{\mu}}\in\sigma_{1}\right\} \in\mathcal{F}_{1}$
and$\left\{ \phi_{1\vec{\mu}}\otimes\phi_{2\vec{\mu}}\in\sigma_{2}\right\} \in\mathcal{F}_{2}$.
\end{thm}
In his contribution, Faris states that a similar result is presented
by Redhead in \cite{Red} at pp.\ 101-102. Redhead claims, it would
have been discovered at first by Suppes and Zanotti \cite{Sup}. The
derivation below will follow closely the lines of Faris: 
\begin{proof}
As shown in section $2.4$, it results from the equivalence property
of Eq.\ (\ref{eq:1abcdefg}) and the passive locality condition of
Eq.\ (\ref{eq:passiveloc}) that 
\begin{equation}
\mathrm{P}_{\phi_{1\vec{\mu}}\otimes\phi_{2\vec{\mu}}}\left(\left.\sigma_{1}\bigcap\sigma_{2}\right|\mathcal{F}_{S}\right)=\mathrm{P}_{\phi_{1\vec{\mu}}\otimes\phi_{2\vec{\mu}}}\left(\left.\sigma_{1}\right|\mathcal{F}_{S}\right)\label{eq:e}
\end{equation}
and, similarly, 
\begin{equation}
\mathrm{P}_{\phi_{1\vec{\mu}}\otimes\phi_{2\vec{\mu}}}\left(\left.\sigma_{1}\bigcap\sigma_{2}\right|\mathcal{F}_{S}\right)=\mathrm{P}_{\phi_{1\vec{\mu}}\otimes\phi_{2\vec{\mu}}}\left(\left.\sigma_{2}\right|\mathcal{F}_{S}\right).\label{eq:f}
\end{equation}
Using the assumption of passive locality again, we get with Eq.\ (\ref{eq:e})
and Eq.\ (\ref{eq:f}): 
\begin{eqnarray}
\mathrm{P}_{\phi_{1\vec{\mu}}\otimes\phi_{2\vec{\mu}}}\left(\left.\sigma_{1}\right|\mathcal{F}_{S}\right) & = & \mathrm{P}_{\phi_{1\vec{\mu}}\otimes\phi_{2\vec{\mu}}}\left(\left.\sigma_{1}\bigcap\sigma_{2}\right|\mathcal{F}_{S}\right)\nonumber \\
 & = & \mathrm{P}_{\phi_{1\vec{\mu}}\otimes\phi_{2\vec{\mu}}}\left(\left.\sigma_{1}\right|\mathcal{F}_{S}\right)\mathrm{P}_{\phi_{1\vec{\mu}}\otimes\phi_{2\vec{\mu}}}\left(\left.\sigma_{2}\right|\mathcal{F}_{S}\right)\nonumber \\
 & = & \left(\mathrm{P}_{\phi_{1\vec{\mu}}\otimes\phi_{2\vec{\mu}}}\left(\left.\sigma_{1}\right|\mathcal{F}_{S}\right)\right)^{2}.\label{eq:j}
\end{eqnarray}

Eq.\ (\ref{eq:j}) implies, that the random variable $\mathrm{P}_{\phi_{1\vec{\mu}}\otimes\phi_{2\vec{\mu}}}\left(\left.\sigma_{1}\right|\mathcal{F}_{S}\right)$
can only have the values $1$ and $0$. The union event $\cup_{i}\omega_{i}$
of outcomes $\omega_{i}$ on the enlarged probability space for which
$\mathrm{P}_{\phi_{1\vec{\mu}}\otimes\phi_{2\vec{\mu}}}\left(\left.\sigma_{1}\right|\mathcal{F}_{S}\right)(\omega_{i})=1$
will be denoted by $\tilde{\mu}_{1S}$ . From this definition and
Eq.\ (\ref{eq:j}), it follows that $\mathrm{P}_{\phi_{1\vec{\mu}}\otimes\phi_{2\vec{\mu}}}\left(\left.\sigma_{1}\right|\mathcal{F}_{S}\right)$
is equal to the indicator function of $\tilde{\mu}_{1S}$: 
\begin{equation}
\mathrm{P}_{\phi_{1\vec{\mu}}\otimes\phi_{2\vec{\mu}}}\left(\left.\sigma_{1}\right|\mathcal{F}_{S}\right)=1_{\tilde{\mu}_{1S}}.\label{eq:indicator}
\end{equation}
According to section $2.1$, $\mathrm{P}_{\phi_{1\vec{\mu}}\otimes\phi_{2\vec{\mu}}}\left(\left.\sigma_{1}\right|\mathcal{F}_{S}\right)$
is measurable with respect to $\mathcal{F}_{S}$. Therefore, we must
have $\tilde{\mu}_{1S}\in\mathcal{F}_{S}$.

Now, we recall the definition 
\begin{equation}
\mathrm{P}_{\phi_{1\vec{\mu}}\otimes\phi_{2\vec{\mu}}}\left(\left.\sigma_{1}\right|\mathcal{F}_{S}\right)\equiv\mathrm{P}\left(\left.\left\{ \phi_{1\vec{\mu}}\otimes\phi_{2\vec{\mu}}\in\sigma_{1}\right\} \right|\mathcal{F}_{S}\right)\label{eq:defabcdefg}
\end{equation}
from section $2.1$. In Eq.\ (\ref{eq:gga}), the unconditional probability
of $\left\{ \phi_{1\vec{\mu}}\otimes\phi_{2\vec{\mu}}\in\sigma_{1}\right\} $
is given by the expectation value 
\begin{eqnarray}
\mathrm{P}_{\phi_{1\vec{\mu}}\otimes\phi_{2\vec{\mu}}}\left(\sigma_{1}\right) & = & \mathrm{EX}\left[\mathrm{P}_{\phi_{1\vec{\mu}}\otimes\phi_{2\vec{\mu}}}\left(\left.\sigma_{1}\right|\mathcal{F}_{S}\right)\right]\\
 & = & \int\mathrm{P}\left(\left.\left\{ \phi_{1\vec{\mu}}\otimes\phi_{2\vec{\mu}}\in\sigma_{1}\right\} \right|\mathcal{F}_{S}\right)\,\mathrm{dP}.\nonumber 
\end{eqnarray}
Similarly, the probability of an event $\tilde{\mu}_{1S}$ on the
enlarged probability space with its measure $\mathrm{P}$ can be computed
by the expectation value of the indicator function $1_{\tilde{\mu}_{1S}}$:
\begin{equation}
\mathrm{EX}\left[1_{\tilde{\mu}_{1S}}\right]=\int1_{\tilde{\mu}_{1S}}\,\mathrm{dP}=\mathrm{P}\left(\tilde{\mu}_{1S}\right).\label{eq:expect12345}
\end{equation}
Using the Eqs.\ (\ref{eq:expectdsadf}), (\ref{eq:defabcdefg}),
(\ref{eq:indicator}), and (\ref{eq:expect12345}) we can conclude
that 
\begin{equation}
\mathrm{P}_{\phi_{1\vec{\mu}}\otimes\phi_{2\vec{\mu}}}\left(\sigma_{1}\right)=\mathrm{P}\left(\tilde{\mu}_{1S}\right).\label{eq:eq2a}
\end{equation}
In section $2.1$, we also have learnt that in case of $\tilde{\mu}_{1S}\neq\varnothing$,
the conditional probability of the intersection between the two events
$\left\{ \phi_{1\vec{\mu}}\otimes\phi_{2\vec{\mu}}\in\sigma_{1}\right\} $
and $\tilde{\mu}_{1S}\in\mathcal{F}_{S}$ is equal to

\begin{eqnarray}
 &  & \mathrm{P}\left(\left.\left\{ \phi_{1\vec{\mu}}\otimes\phi_{2\vec{\mu}}\in\sigma_{1}\right\} \cap\tilde{\mu}_{1S}\right|\mathcal{F}_{S}\right)\\
 &  & \;=\begin{cases}
\mathrm{P}\left(\left.\left\{ \phi_{1\vec{\mu}}\otimes\phi_{2\vec{\mu}}\in\sigma_{1}\right\} \right|\mathcal{F}_{S}\right)\text{ for outcomes in }\tilde{\mu}_{1S},\\
0\text{ for outcomes not in }\tilde{\mu}_{1S}.
\end{cases}\label{eq:defabcdefgh}
\end{eqnarray}
On the other hand, we get with $\tilde{\mu}_{1S}=\varnothing$: 
\begin{equation}
\mathrm{P}\left(\left.\left\{ \phi_{1\vec{\mu}}\otimes\phi_{2\vec{\mu}}\in\sigma_{1}\right\} \bigcap\tilde{\mu}_{1S}\right|\mathcal{F}_{S}\right)=\mathrm{P}\left(\left.\varnothing\right|\mathcal{F}_{S}\right)=0\text{ for all outcomes.}\label{eq:defabcdefgh12}
\end{equation}
The Eqs.\ (\ref{eq:defabcdefgh}), (\ref{eq:defabcdefg}) and (\ref{eq:indicator})
then lead to the relation 
\begin{equation}
\mathrm{P}\left(\left.\left\{ \phi_{1\vec{\mu}}\otimes\phi_{2\vec{\mu}}\in\sigma_{1}\right\} \bigcap\tilde{\mu}_{1S}\right|\mathcal{F}_{S}\right)=\mathrm{P}_{\phi_{1\vec{\mu}}\otimes\phi_{2\vec{\mu}}}\left(\left.\sigma_{1}\right|\mathcal{F}_{S}\right)=1_{\tilde{\mu}_{1S}}.\label{eq:equaility1234}
\end{equation}
Due to Eq.\ (\ref{eq:defabcdefgh12}) and the definition of $\tilde{\mu}_{1S}$,
Eq.\ (\ref{eq:equaility1234}) also holds in case of $\tilde{\mu}_{1S}=\varnothing$.
Computing the expectation value from both sides of Eq.\ (\ref{eq:equaility1234})
yields 
\begin{equation}
\mathrm{P}\left(\left\{ \phi_{1\vec{\mu}}\otimes\phi_{2\vec{\mu}}\in\sigma_{1}\right\} \bigcap\tilde{\mu}_{1S}\right)=\mathrm{P}\left(\tilde{\mu}_{1S}\right)=\mathrm{P}_{\phi_{1\vec{\mu}}\otimes\phi_{2\vec{\mu}}}\left(\sigma_{1}\right).\label{eq:eq2}
\end{equation}

Eqs.\ (\ref{eq:eq2a}) implies that the event $\left\{ \phi_{1\vec{\mu}}\otimes\phi_{2\vec{\mu}}\in\sigma_{1}\right\} \in\mathcal{F}_{1}$
is equivalent to an event $\tilde{\mu}_{1S}\in\mathcal{F}_{S}$. If
an event $A\in\mathcal{F}$ is equivalent to an event $B\in\mathcal{F}$
and $B$ is equivalent to another event $C\in\mathcal{F}$, then $A$
is also equivalent to $C$. The events from $\left\{ \phi_{1\vec{\mu}}\otimes\phi_{2\vec{\mu}}\in\sigma_{1}\right\} \in\mathcal{F}_{1}$,
$\left\{ \phi_{1\vec{\mu}}\otimes\phi_{2\vec{\mu}}\in\sigma_{2}\right\} \in\mathcal{F}_{2}$
and $\tilde{\mu}_{1S}\in\mathcal{F}_{S}$ are in the the sub sigma
algebras $\mathcal{F}_{1}$, $\mathcal{F}_{2}$ and $\mathcal{F}_{S}$
of $\mathcal{F}$ and therefore they are also in $\mathcal{F}$. Furthermore
$\left\{ \phi_{1\vec{\mu}}\otimes\phi_{2\vec{\mu}}\in\sigma_{1}\right\} $
is equivalent to both $\left\{ \phi_{1\vec{\mu}}\otimes\phi_{2\vec{\mu}}\in\sigma_{2}\right\} $and
$\tilde{\mu}_{1S}$. For this reason, the event $\left\{ \phi_{1\vec{\mu}}\otimes\phi_{2\vec{\mu}}\in\sigma_{2}\right\} \in\mathcal{F}_{2}$
is also equivalent to $\tilde{\mu}_{1S}\in\mathcal{F}_{S}$. So we
have the result that passive locality together with the exact anti
correlations of entangled spin states imply that deterministic passive
locality holds. The latter states that the events at the detectors
must be predetermined by another event at preparation stage.
\end{proof}

\section{The relation between Bell's theorem and the Free Will theorem of
Conway and Kochen}

In this section, we review a Proof of Bell's inequalities that is
given by Faris in \cite{Far}. It is interesting, because it relates
Bell's theorem to what Kochen and Conway call the Free Will theorem
\cite{Freewill}.
\begin{thm}
(Faris) If active locality and deterministic passive locality hold,
then Bell's inequality holds.\end{thm}
\begin{proof}
In the previous section, we have shown that the event $\left\{ \phi_{1\vec{A}}\otimes\phi_{2\vec{A}}\in\sigma_{1}=\uparrow\right\} \in\mathcal{F}_{1}$
for an axis $\vec{A}$ set at both detectors 1 and 2 is equivalent
to some event $\tilde{A}_{1S}\equiv\left\{ \chi_{\vec{A}\vec{A}}=1\right\} \in\mathcal{F}_{S}$
that is generated by a random variable which we will denote in the
following as $\chi_{\vec{A}\vec{A}}:=\mathrm{P}_{\phi_{1\vec{A}}\otimes\phi_{2\vec{A}}}\left(\left.\sigma_{1}=\uparrow\right|\mathcal{F}_{S}\right)$:
\begin{eqnarray}
\mathrm{P}_{\phi_{1\vec{A}}\otimes\phi_{2\vec{A}}}\left(\sigma_{1}=\uparrow\right) & = & \mathrm{P}\left(\left\{ \phi_{1\vec{A}}\otimes\phi_{2\vec{A}}\in\left(\sigma_{1}=\uparrow\right)\right\} \right)\nonumber \\
 & = & \mathrm{P}\left(\left\{ \phi_{1\vec{A}}\otimes\phi_{2\vec{A}}\in\left(\sigma_{1}=\uparrow\right)\right\} \bigcap\left\{ \chi_{\vec{A}\vec{A}}=1\right\} \right)\nonumber \\
 & = & \mathrm{P}\left(\left\{ \chi_{\vec{A}\vec{A}}=1\right\} \right)\nonumber \\
 & = & \mathrm{P}\left(\tilde{A}_{1S}\right)
\end{eqnarray}
by active locality, we also can substitute the equivalent random variables
$\phi_{2\vec{B}}$ and $\phi_{2\vec{A}}$ for an event in $\mathcal{F}_{1}$:
\begin{eqnarray}
\mathrm{P}_{\phi_{1\vec{A}}\otimes\phi_{2\vec{B}}}\left(\sigma_{1}=\uparrow\right) & = & \mathrm{P}\left(\left\{ \phi_{1\vec{A}}\otimes\phi_{2\vec{B}}\in\left(\sigma_{1}=\uparrow\right)\right\} \right)\nonumber \\
 & = & \mathrm{P}\left(\left\{ \phi_{1\vec{A}}\otimes\phi_{2\vec{A}}\in\left(\sigma_{1}=\uparrow\right)\right\} \right)\nonumber \\
 & = & \mathrm{P}\left(\left\{ \chi_{\vec{A}\vec{A}}=1\right\} \right)\nonumber \\
 & = & \mathrm{P}\left(\tilde{A}_{1S}\right).\label{eq:eqq-2}
\end{eqnarray}
The event $\tilde{A}_{1S}=\left\{ \chi_{\vec{A}\vec{A}}=1\right\} $
is in $\mathcal{F}_{S}$ and therefore happens at preparation stage.
This preparation stage is in the past light cone of the detectors
1 and 2. If we take the active locality condition seriously, then
the random variables that generate an event in $\mathcal{F}_{S}$
should not depend at all on the pairs of axes that were chosen. Hence
$\chi_{\vec{A}\vec{A}}$ should be equivalent to some other axis independent
random variable $\chi$, or 
\begin{equation}
\chi_{\vec{A}\vec{A}}=\chi
\end{equation}
 and therefore 
\begin{equation}
\left\{ \chi_{\vec{A}\vec{A}}=1\right\} =\left\{ \chi=1\right\} 
\end{equation}
By Eq. (\ref{eq:eqq-2}), a similar reasoning should hold for the
events $\left\{ \phi_{1\vec{A}'}\otimes\phi_{2\vec{B}'}\in\left(\sigma_{1}=\uparrow\right)\right\} \in\mathcal{F}_{1}$that
was measured with an arbitrary pair of axes $\vec{A}',\vec{B}'$ where
$\vec{A}'\neq\vec{A}$ and $\vec{B}'\neq\vec{B}$, and whose equivalent
event in $\mathcal{F}_{S}$ we denote by $A_{1S}=\left\{ \chi_{\vec{A}'\vec{B}'}=1\right\} \in\mathcal{F}_{S}$:
\[
\left\{ \chi_{\vec{A}'\vec{B}'}=1\right\} =\left\{ \chi=1\right\} 
\]
Hence, the event $\tilde{A}_{1S}=\left\{ \chi_{\vec{A}\vec{A}}=1\right\} $
that is associated with some pair of axes $\vec{A},\vec{A}$ at the
detectors 1 and 2 should be equivalent to the event $A_{1S}=\left\{ \chi_{\vec{A}'\vec{B}'}=1\right\} \in\mathcal{F}_{S}$
which is generated by $\chi_{\vec{A}'\vec{B}'}$for an arbitrary pair
of axes $\vec{A}',\vec{B}'$ where $\vec{A}'\neq\vec{A}$ and $\vec{B}'\neq\vec{B}$
or 
\begin{eqnarray*}
\mathrm{P}_{\phi_{1\vec{A}}\otimes\phi_{2\vec{B}}}\left(\sigma_{1}=\uparrow\right) & = & \mathrm{P}\left(\tilde{A}_{1S}\right).\\
 & = & \mathrm{P}\left(\tilde{A}_{1S}\bigcap A_{1S}\right)\\
 & = & \mathrm{P}\left(A_{1S}\right)
\end{eqnarray*}
It is at this point where the conflict with quantum mechanics arises.
The assumption of a spin outcome for an axis $\vec{A}$ which is equivalent
to an event that is equivalent to the same spin outcome measured at
the same detector but with another axis $\vec{A}'\neq\vec{A}$ is
incompatible with quantum mechanics, as we will show below.

A similar argument as given above implies that the event $\left\{ \phi_{1\vec{B}}\otimes\phi_{2\vec{B}}\in\sigma_{2}=\downarrow\right\} \in\mathcal{F}_{2}$
for an axis $\vec{B}$ is equivalent to an event $B_{2S}=\left\{ \chi_{\vec{B}\vec{B}}=1\right\} \in\mathcal{F}_{S}$
and we get with active locality: 
\begin{eqnarray}
\mathrm{P}_{\phi_{1\vec{A}}\otimes\phi_{2\vec{B}}}\left(\sigma_{2}=\downarrow\right) & = & \mathrm{P}\left(\left\{ \phi_{1\vec{A}}\otimes\phi_{2\vec{B}}\in\left(\sigma_{2}=\downarrow\right)\right\} \right)\nonumber \\
 & = & \mathrm{P}\left(\left\{ \phi_{1\vec{B}}\otimes\phi_{2\vec{B}}\in\left(\sigma_{2}=\downarrow\right)\right\} \right)\nonumber \\
 & = & \mathrm{P}\left(\left\{ \phi_{1\vec{B}}\otimes\phi_{2\vec{B}}\in\left(\sigma_{2}=\downarrow\right)\right\} \bigcap\left\{ \chi_{\vec{B}\vec{B}}=1\right\} \right)\nonumber \\
 & = & \mathrm{P}\left(\left\{ \chi_{\vec{B}\vec{B}}=1\right\} \right)\nonumber \\
 & = & \mathrm{P}\left(B_{2S}\right).
\end{eqnarray}
Since $\left\{ \phi_{1\vec{B}}\otimes\phi_{2\vec{B}}\in\left(\sigma_{2}=\downarrow\right)\right\} $
is the complement of $\left\{ \phi_{1\vec{B}}\otimes\phi_{2\vec{B}}\in\left(\sigma_{2}=\uparrow\right)\right\} $,
we can substitute the equivalent events
\begin{eqnarray}
\mathrm{P}_{\phi_{1\vec{A}}\otimes\phi_{2\vec{B}}}\left(\sigma_{1}=\uparrow\bigcap\sigma_{2}=\uparrow\right) & = & \mathrm{P}\left(A_{1S}\bigcap\left(B_{2S}\right)^{c}\right)
\end{eqnarray}
where $\left(B_{2S}\right)^{c}$ denotes the complement of $B_{2S}$ 

We can do this for three axes at the detectors $\vec{A},\vec{B},\vec{C}$.
\begin{eqnarray}
\mathrm{P}_{\phi_{1\vec{A}}\otimes\phi_{2\vec{B}}}\left(\sigma_{1}=\uparrow\bigcap\sigma_{2}=\uparrow\right)+\mathrm{P}_{\phi_{1\vec{B}}\otimes\phi_{2\vec{C}}}\left(\sigma_{1}=\uparrow\bigcap\sigma_{2}=\uparrow\right)\nonumber \\
+\mathrm{P}_{\phi_{1\vec{C}}\otimes\phi_{2\vec{A}}}\left(\sigma_{1}=\uparrow\bigcap\sigma_{2}=\uparrow\right)\nonumber \\
=\mathrm{P}\left(A_{1S}\bigcap\left(B_{2S}\right)^{c}\right)+\mathrm{P}\left(B_{1S}\bigcap\left(C_{2S}\right)^{c}\right)\nonumber \\
+\mathrm{P}\left(C_{1S}\bigcap\left(A_{2S}\right)^{c}\right)\label{eq:noncont1}
\end{eqnarray}
For the event $\left(A_{2S}\right)^{c}\in\mathcal{F}_{S}$ , there
is an equivalent event $\left\{ \phi_{1\vec{A}}\otimes\phi_{2\vec{A}}\in\left(\sigma_{2}=\uparrow\right)\right\} $
at the detectors. Now this event is, by Eq. (\ref{eq:1abcdefg}),
equivalent to $\left\{ \phi_{1\vec{A}}\otimes\phi_{2\vec{A}}\in\left(\sigma_{1}=\downarrow\right)\right\} $,
which, by passive locality, must be equivalent to the event $\left(A_{1S}\right)^{c}\in\mathcal{F}_{S}$.
Therefore $\left(A_{2S}\right)^{c}$ must be equivalent to $\left(A_{1S}\right)^{c}$. 

We have shown above that if active and passive locality hold, the
events at the detectors are equivalent to events that are generated
by setting independent random variables at preparation stage. Hence,
the settings of the measurement devices can not influence the outcomes
at the detectors in a theory that is actively and passively local.
This implies that the equivalence of the events at the two detectors
for an axis $\vec{A}$ can not be destroyed even if at one detector
an axis $\vec{B}$ is chosen. Therefore, we can make the substitutions
of the equivalent events $\left(A_{2S}\right)^{c}$ and $\left(A_{1S}\right)^{c}$,
$\left(B_{2S}\right)^{c}$ and $\left(B_{1S}\right)^{c}$ as well
as $\left(C_{2S}\right)^{c}$and $\left(C_{1S}\right)^{c}$ in Eq.
(\ref{eq:noncont1}), and we arrive at a form of Bell's inequality:
\begin{eqnarray}
\mathrm{P}_{\phi_{1\vec{A}}\otimes\phi_{2\vec{B}}}\left(\sigma_{1}=\uparrow\bigcap\sigma_{2}=\uparrow\right)+\mathrm{P}_{\phi_{1\vec{B}}\otimes\phi_{2\vec{C}}}\left(\sigma_{1}=\uparrow\bigcap\sigma_{2}=\uparrow\right)\nonumber \\
+\mathrm{P}_{\phi_{1\vec{C}}\otimes\phi_{2\vec{A}}}\left(\sigma_{1}=\uparrow\bigcap\sigma_{2}=\uparrow\right)\nonumber \\
=\mathrm{P}\left(A_{1S}\bigcap\left(B_{2S}\right)^{c}\right)+\mathrm{P}\left(B_{1S}\bigcap\left(C_{2S}\right)^{c}\right)\nonumber \\
+\mathrm{P}\left(C_{1S}\bigcap\left(A_{2S}\right)^{c}\right)\nonumber \\
=\mathrm{P}\left(A_{1S}\bigcap\left(B_{1S}\right)^{c}\right)+\mathrm{P}\left(B_{1S}\bigcap\left(C_{1S}\right)^{c}\right)\nonumber \\
+\mathrm{P}\left(C_{1S}\bigcap\left(A_{1S}\right)^{c}\right)\nonumber \\
\leq1\label{eq:noncont2}
\end{eqnarray}
In the last step, we have used that the events whose probabilities
are computed are exclusive. 
\end{proof}
For a theory where the outcomes at the detectors do not depend on
the settings of the measurement devices, the exact anticorrelation,
Eq. (\ref{eq:1abcdefg}), will suffice to derive Bell's inequality.
This is called Bell's first theorem by Faris. 
\begin{defn}
\label{def3}We denote the spin up event at detector 1 for axis $\vec{A}$
as $A_{1}$, for axis $\vec{B}$ as $B_{1}$ and for axis $\vec{C}$
as $C_{1}$. Similarly, the spin down event for detector 2 is denoted
by $A_{2}$ for axis $\vec{A}$, $B_{2}$ for axis $\vec{B}$ and
$C_{2}$ for axis $\vec{C}$. Because of the exact anticorrelations
from quantum mechanics, Eq. (\ref{eq:1abcdefg}), the events $A_{1}$
and $A_{2}$ as well as $B_{1}$and $B_{2}$ and the pair $C_{1}$
and $C_{2}$ are required by definition to be equivalent events with
respect to some probability measure $\mathrm{P}$. Similarly, the
events $A_{1}^{c}$, $A_{2}^{c}$ as well as $B_{1}^{c}$, $B_{2}^{c}$
and the pair $C_{1}^{c}$, $C_{2}^{c}$ are required to be equivalent
events, where the notation $A_{1}^{c}$ denotes the complement event
of $A_{1}$. \end{defn}
\begin{thm}
(Bell's first theorem) Let definition (\ref{def3}) hold. Then, Bell's
inequality holds if we assume that 
\begin{enumerate}
\item All events of the experiment are defined on a single probability space
with a probability measure $\mathrm{P}$.
\item The events $B_{1}^{c}$, $B_{2}^{c}$ are equivalent even in case
we measure an event $A_{1}$ at detector 1 and an event $B_{2}^{c}$
at detector 2. Similarly, the events $C_{1}^{c}$, $C_{2}^{c}$ are
equivalent if we measure an event $B_{1}$ at detector 1 and $C_{2}^{c}$
at detector 2. The events $A_{1}^{c}$, $A_{2}^{c}$ are equivalent
if we measure an event $C_{1}$ at detector 1 and $A{}_{2}^{c}$ at
detector 2.
\end{enumerate}
\end{thm}
\begin{proof}
With the probability measure $\mathrm{P}$, we can write the expression
\[
\mathrm{P}\left(A_{1S}\bigcap\left(B_{2S}\right)^{c}\right)+\mathrm{P}\left(B_{1S}\bigcap\left(C_{2S}\right)^{c}\right)+\mathrm{P}\left(C_{1S}\bigcap\left(A_{2S}\right)^{c}\right)
\]
Since we assumed that the equivalence of the events $A_{1}^{c}$,
$A_{2}^{c}$ and $B_{1}^{c}$, $B_{2}^{c}$ as well as $C_{1}^{c}$
, $C_{2}^{c}$ holds even in case we measure different axes at the
detectors, we can substitute the equivalent events and arrive at Bell's
inequality: 
\begin{eqnarray*}
\mathrm{P}\left(A_{1S}\bigcap\left(B_{2S}\right)^{c}\right)+\mathrm{P}\left(B_{1S}\bigcap\left(C_{2S}\right)^{c}\right)+\mathrm{P}\left(C_{1S}\bigcap\left(A_{2S}\right)^{c}\right)\\
=\mathrm{P}\left(A_{1S}\bigcap\left(B_{1S}\right)^{c}\right)+\mathrm{P}\left(B_{1S}\bigcap\left(C_{1S}\right)^{c}\right)+\mathrm{P}\left(C_{1S}\bigcap\left(A_{1S}\right)^{c}\right)\\
\leq1
\end{eqnarray*}

\end{proof}
It is important to realize that Bell's first theorem does not require
any locality principle to hold in order to derive Bell's inequality.
Instead, Bell's first theorem rests mainly on the assumption that
there would exist events with exact anticorrelations even if different
axes are measured at the separate detectors. However, the measurement
devices could influence the outcomes locally. Thereby the measurement
devices could destroy the exact anticorrelations that are required
in the proof of Bell's first theorem. 

For this reason, Nelson derived Bell's inequality from a model where
the measurement devices are explicitely allowed to alter the outcomes
of the experiment through random variables that depend on the settings
of the detectors. Faris summarizes Nelson's proof of Bell's inequality,
which he calls ``Bell's second theorem'', as follows:
\begin{quote}
The logic of the proof is elementary but subtle, {[}...{]} The first
part of the argument is the observation that when the two magnetic
field gradients are taken in opposite direction, the results exactly
coincide. The deterministic passive locality assumption says that
this implies that the randomness must have been introduced at the
time the particles were prepared, since a later source of randomness
would spoil the coincidence. Thus, with the magnetic field gradient
in opposite direction, each spin result is determined by something
that happened at the preparation stage. The spins may also be measured
in the situation when the magnetic field gradient is not in the opposite
direction to the magnetic field gradient for the other particle. According
to active locality, the results of spin measurements should not depend
on the direction used for the measurement of the other particle. Therefore,
for the magnetic field gradients in arbitrary directions, the spin
results for each particle are still determined by something that happened
at the preparation stage. So, the magnets are not responsible for
the randomness, it must be intrinsic to the system of two particles.
In case of magnetic field gradients that are not the same or opposite,
the dependence is strong but not perfect. This dependence is entirely
due to something that happened at the preparation stage. The magnet
configurations do not affect the probabilities. In this situation,
Bell's first theorem says that the dependence is strong enough so
that there can't be an intrinsic overall outcome that determines the
result of each measurement.
\end{quote}
Hence, Bell's first theorem says that there can be no theory for the
singlet state where the random variables that generate the outcomes
do not depend in some way on the settings of the detectors. On the
other hand, Bell's second theorem says that we have two options if
these random variables do depend on the settings of the measurement
devices: 
\begin{itemize}
\item There may be theories for the singlet state where an event at preparation
stage determines the outcomes at the detectors. In that case, active
locality would have to be be violated. Then, the experimenters could
either 

\begin{itemize}
\item modify the outcomes of a spatially separated measurement station 
\item or the experimenters would not be able to chose their settings freely. 
\end{itemize}
\item On the other hand, if active locality holds, the outcomes of the EPR
experiment have to violate passive locality, which implies that the
outcomes can not be pre-determined by some earlier preparation event. 
\end{itemize}
This result is quite similar to a theorem for spin 1 particles which
was discovered recently by Conway and Kochen, who called it ``Strong
Free-Will Theorem'' \cite{Freewill}. Conway and Kochen assume three
axioms:
\begin{defn}
Axioms of the Strong Free Will theorem
\begin{enumerate}
\item SPIN: The squared spin component of a spin one particle, taken in
three orthogonal directions, is always a permutation of (1,1,0), 
\item TWIN: For entangled spin 1 particles, suppose experimenter A performs
a triple experiment of measuring the squared spin component of particle
a in three orthogonal directions x, y, z, while experimenter B measures
the twinned particle b in one direction, w. Then if w happens to be
in the same direction as one of x, y, z, experimenter B\textquoteright s
measurement will necessarily yield the same answer as the corresponding
measurement by A, 
\item MIN: Assume that the experiments performed by A and B are space-like
separated. Then experimenter B can freely choose any one of the 33
particular directions w, and a\textquoteright s response is independent
of this choice.
\end{enumerate}
\end{defn}
Clearly, the MIN axiom of Conway and Kochen corresponds exactly to
the active locality condition of Nelson. Similarly, the TWIN axiom
corresponds to our Eq. (\ref{eq:1}). However, while the SPIN axiom
holds in quantum mechanics for spin 1 particles, we did not need anything
similar in the proof of Bell's theorem with spin 1/2 particles. Using
SPIN, TWIN and MIN, the ``Strong Free Will Theorem'' of Conway and
Kochen reads in their words\cite{Freewill}. :
\begin{thm}
(Strong Free Will theorem, Conway-Kochen): The axioms SPIN, TWIN and
MIN imply that the response of a spin 1 particle to a triple experiment
is free \textemdash{} that is to say, is not a function of properties
of that part of the universe that is earlier than this response with
respect to any given inertial frame.
\end{thm}
This is exactly what the failure of passive locality implies in probabilistic
terms: If passive locality fails then the outcomes of the experiment
at the detectors are not determined by some earlier event prior to
the measurement process. In that sense, Bell's theorem for spin 1/2
particles can be seen as the analogue of the Strong Free Will theorem
for spin 1 particles.

The author of this note believes that experimenters can choose the
settings of their devices freely. It then follows that theories which
violate active locality also have to violate causality in the sense
of special relativty for individual outcomes. Unfortunately, models
that violate causality can not be written manifestly covariant. Quantum
mechanics can be written in the manifestly covariant form of quantum
field theory. Therefore, the author of this note advocates the view
that models which violate active locality, like Bohmian mechanics
for example, should not be pursued. Instead, the author of this article
thinks that only models where passive locality fails should be considered.

\section{Relations between Bell's original works and the articles of Nelson
and Faris}

\subsection{Bell's first theorem}

In his first article on the hidden variable question \cite{Bell2},
Bell derives his inequality from a set of certain assuptions. We will
show below that Bell's work from \cite{Bell2} corresponds to Bell's
first theorem from section 4. Bell's article starts by defining two
random variables $A(\vec{a},\lambda)=\pm1$ and $B(\vec{b},\lambda)=\pm1$,
where $\vec{a}$ is the axis at detector 1, $\vec{b}$ is the axis
at detector 2 and $\lambda$ is some parameter over which one integrates.
Bell writes:
\begin{quotation}
``The vital assumption is that the result $B$ for particle 2 does
not depend on the setting $\vec{a}$ of the magnet for particle 1
nor $A$ on $\vec{b}$''. 
\end{quotation}
In order to describe a theory with exact anticorrelations, the random
variables are defined to fulfil: 
\begin{equation}
A(\vec{b},\lambda)=-B(\vec{b},\lambda).\label{eq:belleqdsaf}
\end{equation}
Then Bell defines an expectation value:

``If $\rho(\lambda)$ is the probability distribution of $\lambda$,
then the expectation value of the product $\sigma_{1}\vec{a}$ and
$\sigma_{1}\vec{b}$ is 
\begin{equation}
\mathrm{E}=\int d\lambda\rho(\lambda)A(\vec{a},\lambda)B(\vec{b},\lambda)"\label{eq:sfwafsdf}
\end{equation}
Finally, Bell uses Eq. (\ref{eq:belleqdsaf}) and gets 
\begin{equation}
\mathrm{E}=-\int d\lambda\rho(\lambda)A(\vec{a},\lambda)A(\vec{b},\lambda)\label{eq:qdqfddxa}
\end{equation}
and it is as early as here in Bell's first article, where the conflict
with quantum physics.arises. When going from Eq.(\ref{eq:sfwafsdf})
to (\ref{eq:qdqfddxa}), it was implicitely assumed 
\begin{enumerate}
\item That there would be a probability space for all events of the experiment
and one can compute probabilities and expectation values with some
probability measure. 
\item That the exact anticorrelations for outcomes associated with an axis
$\vec{b}$, Eq. (\ref{eq:belleqdsaf}), hold at a detector $A$ in
a case where at the same time, outcomes for an axis $\vec{a}$ are
measured at the same detector $A$. 
\end{enumerate}
The assumptions 1) and 2) were also used in the proof of Bell's first
theorem from section 4. However, in contrast to Bell's own argument,
the proof from section 4 uses assumptions 1) and 2) only. Especially,
in order to derive Bell's inequality, it does not need some hidden
parameter $\lambda$, or random variables $A$ and $B$ with some
kind of locality principle in their definition. Once we assumed that
1) and 2) hold, Bell's inequality followed in section 4 immediately,
with no hidden parameter or locality assumption at all. 

From Bell's first theorem, one can certainly not conclude.that somehow
active locality must fail in quantum mechanics as it is erroneously
done in the book of Dürr \cite{Durr} for example. Instead, Bell's
inequality may fail in nature because the assumption 2) does not hold.
A physicist who wants to elucidate the physical significance of Bell's
first theorem now has to define some set of physical conditions that
lead to the (non-physical) assumptions 2), and that was what we did
in section 4.

\subsection{Bell's second theorem}

In his later article \cite{Bell3}, Bell derived his inequality for
theories which are, in Bell's words ``not deterministic''. By this,
Bell ment that 
\begin{quotation}
``the assigment of values to some beables $\Lambda$ implies, not
necessarily a particular value, but a probability distribution of
some other beable $A$'' 
\end{quotation}
This proof of Bell's inequality is called Bell's second theorem by
Faris\cite{Far}. We will show below that it corresponds to Nelson's
proof in section 2. At first, Bell introduces the notations $\{A|\Lambda\}$
as the 
\begin{quotation}
``probability of a particular value $A$ given the particular value
$\Lambda$''. 
\end{quotation}
Bell then writes: 
\begin{quotation}
``Let $N$ denote a specification of all the beables of some theory
belonging to the overlap of the backward light cone of some spacelike
separated regions 1) and 2)''. Let $A$ be a specification of some
beable localized spacetime region 1 ``and $B$ of some beables in
region 2. Let $\Lambda$ be a specification of some beables from the
remainder of the backward lightcone 1). Then, in a locally causal
theory 
\begin{equation}
\{A|\Lambda,N,B\}=\{A|\Lambda,N\}\label{eq:localcausal}
\end{equation}
whenever both probabilities are given by the theory.

Consider a pair of beables $A$ and $B$, belonging respectively to
regions $1$ and $2$ which happen to have the property $|A|\leq1$
and $|B|\leq1$. Consider the situation in which beables $\Lambda,M,N$
are specified where $N$ is a complete specification of the beables
in the overlap of the lightcones and $\Lambda$and $M$ belong respectively
to the remainders of the two lightcones. 

Consider the joint probability distribution $\{A,B|\Lambda,M,N\}$.
By a standard rule of probability it is equal to $\{A|\Lambda,M,N,B\}\{B|\Lambda,M,N\}$which
by {[}Eq. (\ref{eq:localcausal}){]} is the same as 
\begin{equation}
\{A,B|\Lambda,M,N\}=\{A|\Lambda,M,N,B\}\{B|\Lambda,M,N\}=\{A|\Lambda,N\}\{B|M,N\}\label{eq:Passivesadfs}
\end{equation}
This says simply that correlations between $A$ and $B$ can arise
only because of common causes $N$ .

Consider now the expectiation value of a product 
\[
\mathrm{E}(\Lambda,M,N)=\sum_{A,B}AB\{A|\Lambda,N\}\{B|M,N\}=\overline{A}(\Lambda,N)\overline{B}(M,N)
\]
where $\overline{A}(\Lambda,N)\overline{B}(M,N)$ are functions of
the variables indicated, and $|\overline{A}|\leq1$ and $|\overline{B}|\leq1$
for all values of the arguments.''
\end{quotation}
With Eq. (\ref{eq:Passivesadfs}), Bell has given what Nelson calls
the passive locality condition. This property is denoted as ``locally
causal'' in Bell's work. Furthermore, Bell even has discovered what
this assumption implies: Namely that any correlations in a passively
local theory are due to some common cause. This, however, is something
that is entirely different from the locality principle of special
relativity. Since correlation does not imply causation, a theory may
be completely local in the sense of special relativity, while at the
same time, it describes exact (anti-) correlations at two separated
measurement stations that are not due to some common cause from the
common overlap of the two past lightcones of the stations. 

After a short calculation, Bell then gives his version of the definition
of active locality:
\begin{quotation}
``Suppose now the specifications $\Lambda,M,N$ are given in two
parts $\Lambda=(a,\lambda)$, $M=(b,\mu)$ and $N=(c,\nu)$, where
we are particularly interested in the dependence on $a,b,c$ while
$\lambda,\mu,\nu$ are averaged over some probability distribution,
which may depend on $a,b,c$. In comparison with quantum mechanics,
we will think of $a,b,c$ as variables which specify the experimental
setup in the sense of quantum mechanics, while $\lambda,\mu,\nu$
are in that sense either hidden or irrelevant. Define 
\[
\mathrm{E}(a,b,c)=\overline{\mathrm{E}((a,\lambda),(b,\mu),(c,\nu))}
\]
 where the bar denotes the averaging over $\lambda,\mu,\nu$ just
described. 

\emph{Now applying again the locality hypothesis (\ref{eq:localcausal}),}
the distribution of $\lambda$and $\nu$ must be independent of $b,\mu$,
the latter being outside the relevant backward lightcone. So 
\[
|\mathrm{E}(a,b,c)\pm\mathrm{E}(a,b',c)|\leq\overline{|E((a,\lambda),(b,\mu),(c,\nu))\pm E((a,\lambda),(b',\mu'),(c,\nu))|}
\]
because the mod of the average is less than the average of the mod''.
\end{quotation}
A conditional probability of an event $A$ given an event $B$ on
a probability space with a probability measure $\mathrm{P}$ is something
like 
\[
\mathrm{P}(A|B)=\mathrm{P}(A\cap B)/\mathrm{P}(B)
\]

One can also define some kind of conditional probability with respect
to some collection of events, like a sigma algebra. But the settings
of a measurement device in the EPR experiment are not events, but
parameters. The probabilities of the EPR experiment depend on the
measurement settings. However, they do not depend not on the likelyhood
of these measurement settings, like it would be, if $\left\{ A|(a,\lambda)\right\} $
were ordinary conditional probabilities with respect to the setting
$a$ as some event. 

Hence, Bell's notation of $\left\{ A|(a,\lambda)\right\} $ as ``probability
of a particular value $A$ given the particular values $(a,\lambda)$``
is a highly condensed symbolic notation. It describes a conditional
probability with respect to some set of events $\lambda$ that happened
in the past light cone of $A$ on one hand, and on the other hand
it describes the dependence of an axis dependent probability measure
on the measurement setting as a parameter, like we defined it in Eq.(\ref{eq:def123456}).
Note that the axis dependence of the probability measure on the right
hand side of Eq.(\ref{eq:def123456}) just reflects the axis dependence
of the random variables that generate the outcomes on the left hand
side of Eq.(\ref{eq:def123456}).

Similarly, Bell's definition of a ``locally causal theory'', Eq.
(\ref{eq:localcausal}) is used in two entirely different situations.
One one hand, when Bell writes \emph{``Now applying again the locality
hypothesis''} in \cite{Bell3}, Bell refers to the independence of
the probabilities of certain events at the detectors from the setting
of a remote measurement station, the latter being a parameter. On
the other hand, Bell's definition of a ``locally causal theory''
incorporates, like Nelson's passive locality assumption, the concept
of a common cause of the events at the detectors from an EPR experiment.
However, the assumption of a common cause is entirely unrelated to
the requirements of special relativity. 

In view of Bell's notion of ``local causality'' which is used to
describe two separate conditions, it is no wonder that authors like
Duerr erroneously write in \cite{Durr}:
\begin{quotation}
``Suppose now that \textquotedblleft locality\textquotedblright{}
holds, meaning that the spin measurement on one side has no superluminal
influence on the result of the spin measurement on the other side.
Then we must conclude that the value we predict for the a spin on
R is preexisting. It cannot have been created by the result obtained
on L''
\end{quotation}
Fortunately, Nelson, a mathematician who worked in probability theory,
gave a more formal outline of Bell's locality assumptions. In his
writings \cite{Ne3,Ne4}, Nelson observed that Bell's definition of
a ``locally causal'' theory from \cite{Bell3} in fact consisted
of two separate assumptions, which Nelson called active and passive
locality. Nelson emphasized in his articles that: ``passive locality
is a remnant of deterministic modes of tought'', which is unrelated
to the theory of special relativity. Nelson writes that only the condition
of active locality is connected to the causality requirements from
the theory of special relativity.

The work of Faris \cite{Far} then shows the relation between the
two theorems of Bell. If active and passive locality are both assumed
to hold, then it follows from sections 3 and 4 that the two assumptions
required to show Bell's first theorem are automatically satisfied.
Hence, if both active and passive locality hold, Bell's inequality
also holds. Finally, as we showed in section 4, the assumptions needed
to prove Bell's second theorem can be identified with similar assumptions
that are used in the proof of the Free Will Theorem of Conway and
Kochen for spin 1 particles. This completes the mathematical analysis
of Bell's theorem.

\section{Relation between the assumptions behind Bell's inequality and the
locality principle of Einstein}

As we have seen above, Bell's work basically consists of two theorems.
Bell's first theorem says that there can be no model for the singlet
state where the exact anticorrelations still hold for outcomes that
are associated with an axis $\vec{\mu}$ if a measurement at one detector
is made with an axis $\vec{\nu}\neq\vec{\mu}$. Bell's second theorem
says that there can be no model for the singlet state where both passive
and active locality hold. Active locality is a requirement of special
relativity and the ability of experimenters to chose their axes freely.
On the other hand, passive locality implys that the outcomes at the
detectors are predetermined by some other event that happens in the
overlap of the past cones of the two measurement devices. It is interesting
to ask whether EPR would have called theories that violate active
or passive locality to be complete. In their article \cite{Einstein},
they write:
\begin{quotation}
The following requirement for a complete theory seems to be the necessary
one: every element of the physical reality must have a counterpart
in the physical theory
\end{quotation}
and
\begin{quote}
If, without in any way disturbing a system, we can predict with certainty
the value of a physical quantity, then, there exists an element of
reality corresponding to this physical quantity.
\end{quote}
As we have shown in section $3$, this reality criterion can not be
applied to any theory that violates the passive locality condition
of Eq. (\ref{eq:passiveloc}). One can not predict the outcomes of
a non-deterministic theory. However, in their article, EPR do not
write whether they call a theory where one is unable to predict the
outcomes with certainty, as complete or not. The ``if'' condition
in the above quotation is not equal to ``if and only if''. In fact,
Einstein seemed to notice that the reality condition in \cite{Einstein}
would be problematic, given the non-deterministic nature of quantum
mechanics. In a letter to Schroedinger, \cite{How}, Einstein complained
that the main point which he wanted to make would have been ``buried
by erudition''. Einstein then wrote own articles on his views about
quantum mechanics. In his own writings, the reality criterion becomes
a quite different one. He writes 1948 in a letter to Born: 
\begin{quote}
I just want to explain what I mean when I say that we should try to
hold on to physical reality. We are, to be sure, all of us aware of
the situation regarding what will turn out to be the basic foundational
concepts in physics: the point-mass or the particle is surely not
among them; the field, in the Faraday-Maxwell sense, might be, but
not with certainty. But that which we conceive as existing (real)
should somehow be localized in time and space. That is, the real in
one part of space, A, should (in theory) somehow exist independently
of that which is thought of as real in another part of space, B. If
a physical system stretches over the parts of space A and B, then
what is present in B should somehow have an existence independent
of what is present in A. What is actually present in B should thus
not depend upon the type of measurement carried out in the part of
space, A; it should also be independent of whether or not, after all,
a measurement is made in A. If one adheres to this program, then one
can hardly view the quantum-theoretical description as a complete
representation of the physically real. If one attempts, nevertheless,
so to view it, then one must assume that the physically real in B
undergoes a sudden change because of a measurement in A. 
\end{quote}
Clearly, the notion of ``independence'' here is not in the sense
of statistical independence, or the ``conditional independence''
that is described by the passive locality condition of Eq. (\ref{eq:passiveloc}).
What Einstein meant is more in the sense of Nelson's active locality
conditions of Eqs. (\ref{eq:activeloc}) and (\ref{eq:activeloc2-1}):
\begin{equation}
\phi_{1\vec{\mu}}=\hat{\phi}_{1\vec{\mu}},\label{eq:activelocaaaa}
\end{equation}
which says that the random variable $\phi_{1\vec{\mu}}$ that describes
the outcome at detector 1 with an axis $\vec{\mu}$ in case detector
2 is at an axis $\vec{\nu}$, has the same outcome as as the random
variable $\hat{\phi}_{1\vec{\mu}}$ that describes the result at detector
1 with axis $\vec{\mu}$ in case detector 2 has an axis $\vec{\nu}'\neq\vec{\nu}$.
This active locaity condition is the mathematical definition which
ensures that an outcome at one detector can not be modified by the
choice of some setting at another detector if the latter is spacelike
separated. 

Einstein wrote the letter quoted above in 1948. It took time until
1986, when Edward Nelson explicitly showed that the locality principle
which Einstein wanted to implement into quantum mechanics, was not
forbidden by Bell's theorem for a random variable model of the singlet
state. In the words of a recent article of Nelson \cite{Ne8}: 
\begin{quote}
In quantum mechanics, if there are two dynamically uncoupled systems,
an alteration of the second system in no way affects the first, even
if the two systems are entangled. 
\end{quote}
However, in order to maintain locality in the sense of special relativty
as well as the assumption of free will of the experimenters, any theory
that describes quantum mechanics must give up determinism, since latter
is a consequence of passive locality. In the words of Faris:
\begin{quote}
A violation of active locality would be upsetting, since it would
mean that an active intervention at one point could influence the
outcome at distant points. However, a violation of passive locality
would only mean that dependence between simultaneous events at distant
locations need have no explanation in terms of prior events. This
is not clearly ruled out, but it is not evident how to construct such
a theory. 
\end{quote}
Nelson's analysis of Bell's theorem also shows that unfortunately,
quantum mechanics contains no principle which definitely forbids non-local
theories. Quantum mechanics itself just defines some kind of locality
principle for the probabilities of events through equations like:
\begin{equation}
\mathrm{P}_{\vec{\mu}\vec{\nu}}\left(\sigma_{1\vec{\mu}}\right)=\mathrm{P}_{\vec{\mu}\vec{\nu}'}\left(\sigma_{1\vec{\mu}}\right)\label{eq:nonsignallinga}
\end{equation}
with an axis dependent probability measure $\mathrm{P}_{\vec{\mu}\vec{\nu}}$
and $\vec{\nu}\neq\vec{\nu}'$.. Therefore, quantum mechanics does
not say anything on whether the active locality condition holds for
individual outcomes or not. One is always free to reproduce the quantum
mechanical probabilities with a model that violates active locality,
like Bohmian mechanics, or Nelson's stochastic mechanics. This is
not what one should expect from a physical theory like quantum mechanics
that can be written manifestly covariant, and it is the reason for
the interpretational problems of quantum mechanics.

In contrast to Eq. (\ref{eq:nonsignallinga}), the active locality
condition of Eq. (\ref{eq:activelocaaaa}) is a relation that holds
for random variables which describe individual outcomes. So one first
has to find a model with appropriate random variables if one wants
to implement Nelson's active locality condition into quantum mechanics.
Only if we have such a random variable model, we are able to exclude
the various causality violating, and therefore non-physical interpretations
of quantum theory, like Bohmian mechanics. From Einstein's quotation
above, one can assume that this was the thing that Einstein was concerned
about.

\end{document}